\documentclass[nofootinbib,amsfonts,prd,aps]{revtex4}
\usepackage{subfig}
\usepackage{amsmath,amssymb,amsfonts,amsthm,mathrsfs}
\usepackage{stackrel}
\usepackage{graphicx}
\usepackage{nameref}

\begin{document}

\title{Quantum fluctuating geometries and the information paradox II}

\author{Rodrigo Eyheralde$^{1}$, Rodolfo Gambini$^{1}$, 
Jorge Pullin$^{2}$}
\affiliation {
1. Instituto de F\'{\i}sica, Facultad de Ciencias, 
Igu\'a 4225, esq. Mataojo, 11400 Montevideo, Uruguay. \\
2. Department of Physics and Astronomy, Louisiana State University,
Baton Rouge, LA 70803-4001}

\begin{abstract}
  In a previous paper we discussed corrections to Hawking radiation
  from a collapsing shell due to quantum fluctuations of the shell and
  the resulting horizon. For the computation of the quantum
  corrections we used several approximations. In this paper
  we take into account effects that were neglected in the previous
  one. We find important corrections including non-thermal
  contributions to the radiation at high frequencies and a frequency
  dependent time scale at which the emission of thermal radiation of frequency $\omega$
  cuts off. Such scale tends to infinity in the limit of a classical shell. The
  fact that one has almost from the outset non-thermal radiation
  has significant implications for the information paradox. In particular the
  amount of non-thermality is considerably larger than what we had
  estimated before. A naive estimate of
  the evaporation time leads to a much faster evaporation than in the
  usual Hawking analysis. 
\end{abstract}

\maketitle

\section{Introduction} 

Hawking radiation has been studied for a collapsing shell going all
the way back to Boulware in 1976 \cite{Boulware}. Most studies have treated the shell
as a classical collapsing object. In a previous paper \cite{previous} we have studied
the Hawking radiation produced by a collapsing quantum shell. We did
it in the geometric optics approximation. In it, one
considers ingoing and outgoing light rays and how they relate to each
other via the parameters of the shell, namely its ADM mass and the
position at past infinity from which the shell is launched. When the
ADM mass and the position are turned into quantum operators acting on
a Hilbert space for the geometry created by the shell, so do the
relations between ingoing and outgoing rays. In the standard geometric
optic treatment of Hawking radiation, the relations are used to
construct the Bogoliubov coefficients. In our case the latter become
quantum operators acting on the Hilbert space of the geometry. We
found that the profile of the Hawking radiation as a function of time
contains information about the initial state of the collapsing quantum
shell. In particular, certain correlations of the Hawking radiation
that vanish in the classical case, are non-vanishing in the quantum
one.  Since the Hawking radiation for a large black hole occurs
entirely in a low-curvature region of space-time, our calculation
shows that non-trivial quantum effects can occur in such types of
regions. Our calculations involved several approximations, which we
study in greater detail in the current paper. We find important
corrections. The main message is that, even at rather short times, the
radiation becomes non-thermal.  The emission of the total mass of the
black hole is faster than in the case of traditional Hawking
radiation. In section 2 we summarize the previous results, showing that
they correspond to a naive semi-classical limit that omits certain
important quantum effects. Section 3 carries out the full quantum
calculation which we characterize in terms of effective c-number Bogoliubov
coefficients. In section 4 we compute the number of particles emitted,
showing the non-thermal nature of their spectrum and bounds for the
total the total thermal energy emitted.
We end with a discussion.

\section{General framework}
\subsection{Summary of previous results}

The computation of Hawking radiation using the geometric optics
approximation has a long history going back all the way to Hawking's
original 1975 calculation \cite{h74}. Boulware was the first to
consider the radiation of a collapsing null shell \cite{Boulware}. The metric of a collapsing shell is given by,
\begin{equation}
ds^{2}=-\left(1-\frac{2M\theta(v-v_{s})}{r}\right)dv^{2}+2dvdr+r^{2}d\Omega^{2},
\end{equation}
where $v_s$ represents the position of the shell (in ingoing
Eddington--Finkelstein coordinates) and $M$ its mass. We are using
units where $G=c=1$  Its associated Penrose diagram is given in figure 1. 
  \begin{figure}
\includegraphics[height=8.5cm]{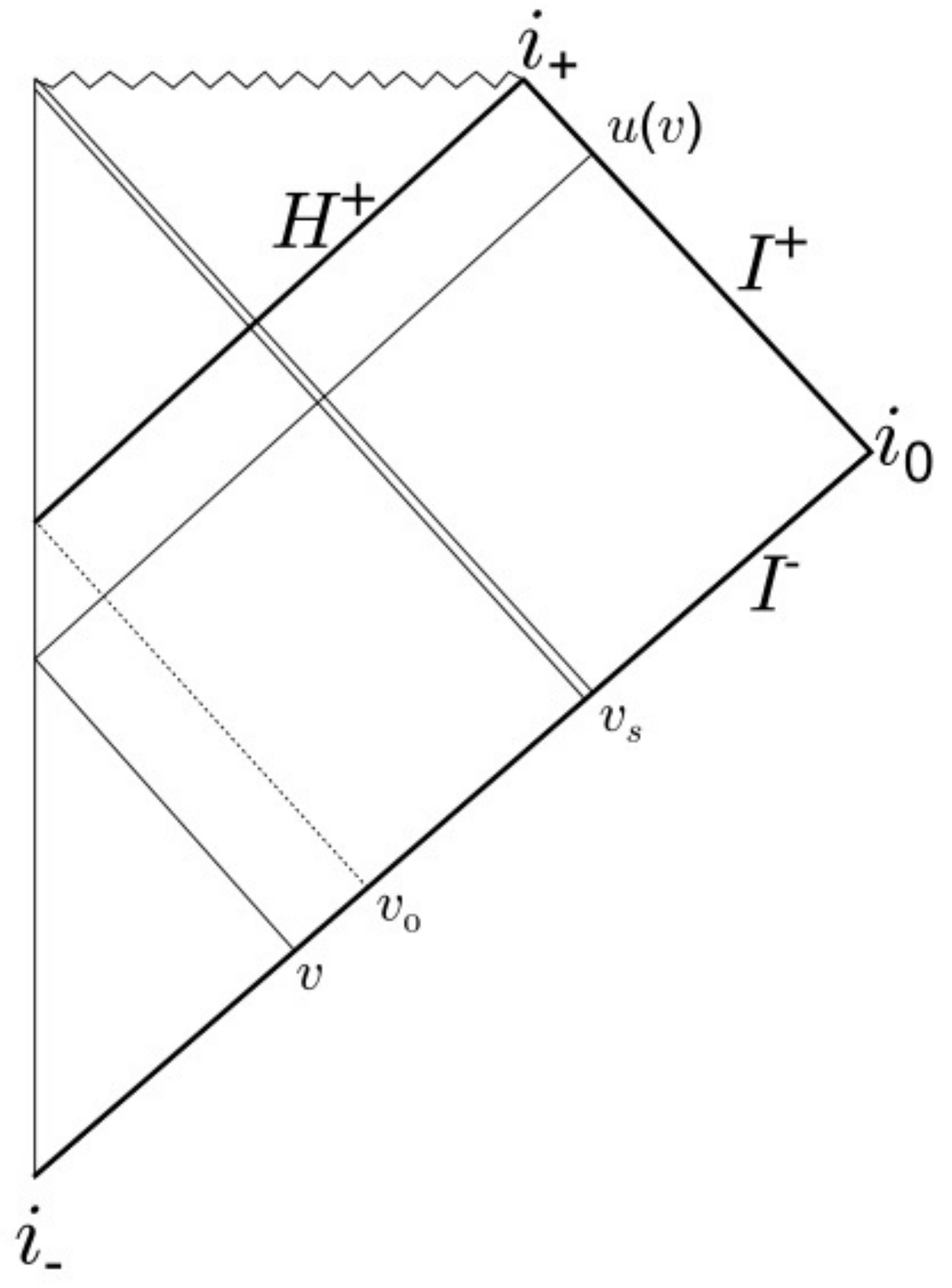}
\caption{The Penrose diagram of collapsing shell. $v_s$ is the
  position at $I^-$ from which the shell is sent in. Light rays sent
  in to the left of $v_0$ make it to $I^+$, and rays sent in to
  the right of $v_0$ get trapped in the black hole.}
\end{figure}

To use the geometric optics approximation one considers light rays that leave $I^-$ 
with coordinate $v$ less than 
$v_{0}=v_{s}-4M$ and escape to $I^{+}$ with the rest trapped in the
black hole that forms. They reach $I^+$ with a coordinate,
\begin{equation}
u(v)=v-4M \ln \left(\frac{v_{0}-v}{4M_{0}}\right)\label{eq:u(v)},
\end{equation}
where $M_0$ is an arbitrary parameter that is usually chosen as
$M_0=M$. In our case, since we are considering a  quantum black hole we will
take $M_0$ to be the mean value of the mass. $M_0$ is related with the
definition of the tortoise coordinate $u$, which
involves a constant of integration. One uses this identity to relate the ``in'' modes of the scalar field at $I^-$,
\[
\psi_{lm\omega'}(r,v,\theta,\phi)=\frac{e^{-i\omega'v}}{4\pi r\sqrt{\omega'}}Y_{lm}(\theta,\phi),
\]
with the ``out'' modes at $I^+$,
\[
\chi_{lm\omega}(r,u,\theta,\phi)=\frac{e^{-i\omega u}}{4\pi r\sqrt{\omega}}Y_{lm}(\theta,\phi),
\]
to compute the Bogoliubov coefficients,
\[
\alpha_{\omega\omega'}=\left\langle \chi_{lm\omega},\psi_{lm\omega'}\right\rangle, 
\]
\[
\beta_{\omega\omega'}=-\left\langle \chi_{lm\omega},\psi_{lm\omega'}^{*}\right\rangle. 
\]

That was the summary of the calculation of Hawking radiation on the background of a classical collapsing shell in the geometric optics approximation. To consider the case of quantum collapsing shells we recall that the ADM mass $M$ and the position at $I^-$ from which the shell is sent in are  a complete set of Dirac observables and canonically conjugate to each other \cite{lwf,previous}. One can promote them to quantum operators with commutators,
\begin{equation}
\left[\widehat{M},\widehat{v}_{s}\right]=i\hbar\widehat{I}\label{eq:conmutacio_M_v},
\end{equation}
with $\widehat{I}$ the identity operator. It is actually more convenient to use $v_0=v_s -4M$ instead of $v_s$. In terms of the quantum operators one can write a quantum operatorial relationship between the operator associated with the ingoing position in $I^-$ of a light ray and the outgoing position at $I^+$,
\begin{equation}
\hat{u}\left(v,\widehat{v}_{0},\widehat{M}\right)=v\widehat{I}-2\left[\widehat{M}\ln\left(\frac{\widehat{v}_{0}-v\widehat{I}}{4M_{0}}\right)+\ln\left(\frac{\widehat{v}_{0}-v\widehat{I}}{4M_{0}}\right)\widehat{M}\right]\label{eq:operador_u}.
\end{equation}
The operators act on states of the quantum geometry, which in the mass
representation are given by $\varphi(M)$. With the above operators one
can now promote the Bogoliubov coefficients to operators acting on the
states of the quantum geometry,
\begin{equation}
\hat{\beta}_{\omega\omega'}=-\frac{1}{2\pi}\sqrt{\frac{\omega'}{\omega}}\underset{\delta\to0}{\lim}{\int_{-\infty}^{+\infty}}dv\theta\left(\widehat{v}_{0}-v\widehat{I}\right)\exp\left({-i\omega\hat{u}_{\delta}(v)- i\omega'v}\right)\theta\left(\widehat{v}_{0}-v\widehat{I}\right)\label{eq:coef_bogol_cuanticos}.
\end{equation}
The above computation requires the extension of the operator $\hat{u}$
to include all the range of rays that start at $I^-$ including those
that would fall into the black hole.
Details can be seen in our previous paper \cite{previous}. The result
is the operator $\hat{u}_\delta$. We need to solve its eigenvalue
problem. It turns out that the spectrum of $\hat{u}_\delta$ is
degenerate with degeneracy two. 
This leads us to choose two
independent eigenstates of $\hat{u}$, 
\begin{equation}
\psi_{u}^{1}(x)=\left\{ \begin{array}{l}
\frac{1}{\sqrt{8\pi\hbar\left|\ln(\delta)\right|}}\exp\left(\frac{iM_{0}}{\hbar}(u-v)\frac{x-\delta}{\ln(\delta)}\right),\quad x<\delta\\
\frac{1}{\sqrt{8\pi\hbar\left|\ln(x)\right|}}\exp\left(\frac{iM_{0}}{\hbar}(u-v)\left[{\rm
      li}\left(x\right)-{\rm li}\left(\delta\right)\right]\right),\quad\delta\leq x<1\\
0,\quad x\geq 1
\end{array}\right.\label{eq:autoestados_de_u_1}
\end{equation}
\begin{equation}
\psi_{u}^{2}(x)=\left\{ \begin{array}{l}
0,\quad x\leq 1\\
\frac{1}{\sqrt{8\pi\hbar\left|\ln(x)\right|}}\exp\left(\frac{iM_{0}}{\hbar}(u-v)\left[{\rm
      li}\left(x\right)-{\rm li}\left(\delta\right)\right]\right),\quad x>1
\end{array}\right.\label{eq:autoestados_de_u_2}
\end{equation}
where $x=({v_{0}-v})/({4M_{0}})$
and we have chosen them as orthonormal. We adopt the notation 
$\left|u,J\right\rangle _{\delta}$ with $J=1,2$ for these states.

With this we can compute the expectation values of the Bogoliubov
coefficients for different states of the quantum geometry,
\[
\left\langle
  \hat{\beta}\right\rangle_{\omega\omega'}=-\frac{1}{2\pi}\sqrt{\frac{\omega'}{\omega}}\underset{\delta\to0}{\lim}\left\langle
  \Phi\right|
{\int_{-\infty}^{+\infty}}dv{\int_{-\infty}^{+\infty}}dv_{0}\left|v_{0}\right\rangle \left\langle v_{0}\right|\theta\left(\widehat{v}_{0}-v\widehat{I}\right)e^{-i\omega\hat{u}_{\delta}(v)- i\omega'v}\times
\]
\[
\times\underset{J=1,2}{\sum}{\int_{-\infty}^{+\infty}}du\left|u,J\right\rangle _{\delta }\left\langle u,J\right|_{\delta} {\int_{-\infty}^{+\infty}}dv'_{0}\left|v'_{0}\right\rangle \left\langle v'_{0}\right|\theta\left(\widehat{v}_{0}-v\widehat{I}\right)\left|\Phi\right\rangle.
\]

We also found an expression for the expectation value of the density matrix of a scalar field on the background of a quantum shell,
\[
\left\langle  \rho_{\omega_1\omega_2}^{QS}\right\rangle \equiv
\left\langle\Phi\right|\left\langle 0_{in}\right|\hat{a}_{\omega_1}^{\dagger}(\hat{M},\hat{v}_0)\hat{a}_{\omega_2}(\hat{M},\hat{v}_0)\left|0_{in}\right\rangle\left|\Phi\right\rangle
=\int_{0}^{\infty} d\omega'\left\langle\hat{\beta}_{\omega_1\omega'}\hat{\beta}_{\omega_2\omega'}^{*}\right\rangle,
\] 
where $\vert \Phi\rangle$ is the quantum state of the shell centered
in given values $\bar{M}, \bar{v_0}$ of the ADM mass and the position
along $I^-$ of the last light ray that escapes to $I^+$ (as we
mentioned, it is equivalent to use this quantity in lieu of the position of the shell at $I^-$). In the above expression  $\hat{a}, \hat{a}^\dagger$ are the annihilation and creation operators of the quantum field of the Hawking radiation (we take it to be a scalar field for simplicity)
and  $\left|0_{in}\right\rangle$ is the vacuum of the Hawking radiation.

We take a state for the shell that in the $v_0$ representation can be written as,
\begin{equation}
\psi\left(v_{0}\right)\equiv\left<v_0\right|\left.\varphi\right>=\frac{1}{\sqrt{2\pi\hbar}}\int dM\varphi\left(M\right)\exp\left({-i\frac{M \left[v_0-\bar{v}_0\right]}{\hbar}}\right)\label{fourier}
\end{equation}
with  $\varphi(M)$ 
a complex function centered in $\bar{M}$ that satisfies $\int dM\left|\varphi\left(M\right)\right|^2=1$. 

The result found in our previous paper \cite{previous} (equation above (39)) for the
expectation value density matrix of the Hawking radiation for a
quantum shell is,
\[
\left\langle  \rho_{\omega_1\omega_2}^{QS}\right\rangle=\frac{\left(2M_0\right)^2}{\pi^2}\int_{0}^{\infty} d\omega'\frac{\omega'}{\sqrt{\omega_1\omega_2}}
{\int}_{0}^{+\infty}dxdx'\exp\left({-i4M_0\left[\omega'+\omega_2\right]x'}\right)\exp\left({i4M_0\left[\omega'+\omega_1\right]x}\right)
\times
\]
\[
\times \exp\left({i4M_0\left[\omega'+\bar{\omega}\right]\Delta_{\omega_1\omega_2}(x,x')}\right)
\sqrt{\frac{\left|\ln(\bar{x}_{\omega_1}(x))\right|\left|\ln(\bar{x}_{\omega_2}(x'))\right|}{\left|\ln(x)\right|\left|\ln(x')\right|}}\] \[\times{{\int}_{-\infty}^{+\infty}}ds e^{i\Delta\omega s}\psi^{*}(s-2M_0\Delta_{\omega_1\omega_2}(x,x'))\psi(s+2M_0\Delta_{\omega_1\omega_2}(x,x')),
\]
where $\Delta_{\omega_1\omega_2}(x,x')=\Delta_{\omega_2}(x')-\Delta_{\omega_1}(x)$,
$\Delta_\omega(x)=x-\bar{x}_\omega(x)=x-{\rm li}^{-1}\left({\rm
    li}(x)-\delta_\omega\right)$, $\bar{x}_{\omega_i}(x)={\rm li}^{-1}\left[{\rm li}(x)-\omega_i 
  \hbar/M_0\right]$,  ${\rm li}$ is the logarithmic integral, 
$\delta_\omega=\frac{\hbar\omega}{M_0}$ ,
$\Delta\omega=\omega_2-\omega_1$ and
$\bar{\omega}=\frac{\omega_1+\omega_2}{2}$.

This can be rewritten as,
\begin{equation}
\left\langle  \rho_{\omega_1\omega_2}^{QS}\right\rangle=\int dM \varphi^*\left(M-\frac{\Delta\omega\hbar}{2}\right)\varphi\left(M+\frac{\Delta\omega\hbar}{2}\right)\int_{0}^{\infty} d\omega'\beta^{QS}_{\omega_1\omega'}(M,\bar{\omega})\left[\beta^{QS}_{\omega_2\omega'}(M,\bar{\omega})\right]^*,\label{eq:rho_QS1}
\end{equation}
in terms of the c-number quantity,
\begin{eqnarray}
\beta^{QS}_{\omega\omega'}(M,\bar{\omega})&=&-\frac{2M_0 e^{-i\left[\omega+\omega'\right]\bar{v}_0}}{\pi}\sqrt{\frac{\omega'}{\omega}}
\nonumber\\
&&\times\lim_{\epsilon\to0}
\int_{0}^{+\infty}dx\exp\left({-\epsilon x}\right)\exp\left(i4M_0\left[\omega'+\omega\right] x\right)
\sqrt{\frac{\left|\ln(\bar{x}_{\omega}(x))\right|}{\left|\ln(x)\right|}} \exp\left({i4\omega m(\omega',\bar{\omega})\frac{\Delta_{\omega}(x)}{\delta_\omega}}\right),
\label{eq:beta_varx}
\end{eqnarray}
that plays the role of an effective Bogoliubov coefficient, with
$m(\omega',\bar{\omega})=M-\left(\omega'+\bar{\omega}\right)\hbar$. 
The regulator $\epsilon$ is introduced in order to make the integral convergent since we are in a basis of plane waves.
% \subsection{Limitations of the model} 
% An obvious limitation of the model is that the effects of the radiation on the black hole and on the amount of energy available to be radiated are neglected (backreaction). To take into account the finite amount of energy available to be radiated we will only consider the contribution to the radiation of modes with frequencies 
% $\omega<\omega_{BH}$ with 
% \[\hbar\omega_{BH}\sim k \bar{M}\] with $k$ smaller than one and with  $\bar{M}$
% the mass of the black hole. In addition one cannot have frequencies of
% the order of Planck's frequency $\hbar \omega_{P}\sim M_{P}$ with 
% ${M}_{P}$ Planck's mass. Higher frequencies would lead to
% quantum gravity effects and our analysis would not be valid.
% It is worthwhile reminding that Hawking radiation, since it is thermal, has a reference frequency 
% ($\omega_0$) at which the peak of emissions occur. 
% It is given by,
% \[8\pi\bar{M}\omega_0\sim 2.82144.\]
% For a macroscopic black hole $\omega_{BH}\gg \omega_0$ and therefore,
% \[2,82144\ll 8\pi\bar{M}\omega_{BH}\sim8\pi k
%   \bar{M}\frac{\bar{M}}{\hbar}=8\pi k
%   \left(\frac{\bar{M}}{M_P}\right).\]
% This is not compatible with Planck mass black holes, exhibiting
% another limitation of the model.

\subsection{Semiclassical approximation: naive version of corrections
  to the limit  $\hbar\to0$}\label{sec:lim_ingenuo}
Expression (\ref{eq:rho_QS1}) has the complete information of the
geometric optics approximation for Hawking radiation on the background
of a quantum shell and should therefore include the usual results for
Hawking radiation from a classical collapsing shell when $\hbar\to0$
(the shell variables become classical
but the radiation is kept quantum, otherwise it vanishes). To take 
such a limit is to set $\hbar\to0$ in the integrand of
(\ref{eq:beta_varx}).One gets
\[\beta^{QS}_{\omega\omega'}(M,\bar{\omega})\to\beta^{CS}_{\omega\omega'}(M)=-\frac{2M_0e^{-i\left[\omega+\omega'\right]\bar{v}_0}}{\pi}\sqrt{\frac{\omega'}{\omega}}\lim_{\epsilon\to0}
{\int}_{0}^{+\infty}dxe^{-\epsilon x}\exp\left({i4M_0\left(\omega'+\omega\right)x}\right) \exp\left({i4M\omega\ln(x)}\right)\]
that agrees with the standard result for Hawking radiation for
classical shells (CS). 

In our previous paper \cite{previous} we considered an approximation
in which we kept the states of the quantum geometry but took the limit
$\hbar\to 0$ in the Bogoliubov coefficients in the manner
discussed. We will call this approximation the ``naive limit''. We did this in the
expectation that the corrections this approximation neglected were small. We will see in this paper
that they are not. 

To evaluate the expectation value of the density matrix in the naive
limit, we start with the change of variable $y=\ln(x)$. The above
expression becomes
\begin{eqnarray}\beta^{CS}_{\omega\omega'}(M)&=&-\frac{2M_0e^{-i\left[\omega+\omega'\right]\bar{v}_0}}{\pi}\sqrt{\frac{\omega'}{\omega}}\lim_{\epsilon\to0}
{\int}_{-\infty}^{+\infty}dy\exp\left({-\left[\epsilon-i4M_0\left(\omega'+\omega\right)\right]e^y}\right) \exp\left({(1+i4M\omega)y}\right)\nonumber\\
&=&-\frac{2M_0e^{-i\left[\omega+\omega'\right]\bar{v}_0}}{\pi}\sqrt{\frac{\omega'}{\omega}}\lim_{\epsilon\to0}\left[
{\int}_{0}^{+\infty}dy\exp\left({-\left[\epsilon-i4M_0\left(\omega'+\omega\right)\right]e^{-y}}\right) \exp\left({-(1+i4M\omega)y}\right)\right.\nonumber\\
&&\left.+{\int}_{0}^{+\infty}dy\exp\left({-\left[\epsilon-i4M_0\left(\omega'+\omega\right)\right]e^y}\right)
\exp\left({(1+i4M\omega)y}\right)\right].
\end{eqnarray}
 Using the identities for the upper and lower incomplete Gamma functions,
\begin{equation}
\Gamma(a,z)=z^a\int_0^{+\infty}\exp\left(at-ze^t\right)dt,\quad {\rm Re(z)}>0\label{eq:Gama_Mayus},
\end{equation}
\begin{equation}
\gamma(a,z)=z^a\int_0^{+\infty}\exp\left(-at-ze^{-t}\right)dt,\quad {\rm Re(a)}>0\label{eq:Gama_minus},
\end{equation}
\[\Gamma(a)=\gamma(a,z)+\Gamma(a,z).\]
we get
\begin{eqnarray}
\beta^{CS}_{\omega\omega'}(M)&=&-\frac{2M_0e^{-i\left[\omega+\omega'\right]\bar{v}_0}}{\pi}\sqrt{\frac{\omega'}{\omega}}\lim_{\epsilon\to0}
\exp\left({-(1+i4M\omega)\ln\left[\epsilon-i4M_0\left(\omega'+\omega\right)\right])}\right)\nonumber\\
&&\times\left[\gamma\left(1+i4M\omega,\epsilon-i4M_0
\left[\omega'+\omega\right]\right)+\Gamma\left(1+i4M\omega,\epsilon-i4M_0\left[\omega'+\omega\right]\right)\right]\nonumber\\
&=&\frac{-i\exp\left({-i\left[\omega+\omega'\right]\bar{v}_0}\right) e^{-2M\omega\pi}}
{2\pi\sqrt{\omega}}\frac{\omega'}{\omega'+\omega}
\frac{\exp\left(-i4M\omega\ln\left(4M_0\left[\omega'+\omega\right]\right) \right)}
{\sqrt{\omega'}}
\Gamma\left(1+i4M\omega\right)\label{eq:bogol_CS}.
\end{eqnarray}
 This is the expression of the Bogoliubov coefficient $\beta$ that we
 obtained in the previous paper and that includes Hawking radiation in
 the  long time limit but includes non-thermal corrections for early
 times, as one expects for the radiation of a collapsing shell.
Substituting in (\ref{eq:rho_QS1}) we get
\begin{eqnarray}
\left\langle  \rho_{\omega_1\omega_2}^{QS}\right\rangle_{\rm Naive\, limit}&=&\int dM \varphi^*\left(M-\frac{\Delta\omega\hbar}{2}\right)\varphi\left(M+\frac{\Delta\omega\hbar}{2}\right)\frac{8M^2\sqrt{\omega_1\omega_2}
\exp\left({i\Delta\omega\bar{v}_0}\right)}{\pi e^{4M\bar{\omega}\pi}}\Gamma\left(i4M\omega_1\right)\Gamma\left(-i4M\omega_2\right)\nonumber\\
&&\times\frac{1}{2\pi}\int_{0}^{+\infty} \frac{d\omega'}{\omega'}\frac{(\omega')^2}{(\omega'+\omega_1)(\omega'+\omega_2)}
\exp\left({i4M\left[\omega_2\ln\left(4M_0\left[\omega'+\omega_2\right]\right)
-\omega_1\ln\left(4M_0\left[\omega'+\omega_1\right]\right)\right]}\right).\label{15}
\end{eqnarray}
Notice that we have taken $\beta_{\omega \omega'}$ to be classical but
kept the $\hbar$ dependence in the quantum states. This was the
approximation we used in our previous paper and only partially
captures the departures from thermality of the distribution of
radiated energies. In this paper we will develop a better
approximation and we will see significantly different behavior.

\subsection{Computation of the radiated energy in the naive limit}

In order to compare with the result we will obtain in this paper we
need an expression for the amount of energy radiated as Hawking
radiation in this naive limit. The radiated energy can be obtained from the
diagonal terms of (\ref{15}), that is, the number of particles per
unit frequency. From there, the time of evaporation of the black hole,
assuming the radiation maintains the same form (i.e. ignoring
backreaction) can be estimated.

To compute the integral, we start by rewriting the Gamma function,
\begin{equation}
\Gamma(ix)=\sqrt{\frac{\pi}{x\sinh(\pi x)}}e^{i\phi(x)},
\label{eq:gamma_ix1}
\end{equation}
with $\phi(x)={\rm Im}\left[\ln\left(\Gamma(ix)\right)\right]$ and
carry out the change of variable,
\begin{equation}t=\frac{4M}{\Delta\omega}\left[\omega_2\ln\left(4M_0\left[\omega'+\omega_2\right]\right)-\omega_1\ln\left(4M_0\left[\omega'+\omega_1\right]\right)\right], \label{t}
\end{equation}
and recalling that $x=4 M\omega_1$ or $x=-4M \omega_2$, this leads to,
\begin{equation}
\left\langle \rho_{\omega_1\omega_2}^{QS}\right\rangle_{\rm Naive\, limit}=\int dM \varphi^*\left(M-\frac{\Delta\omega\hbar}{2}\right)\varphi\left(M+\frac{\Delta\omega\hbar}{2}\right)\frac{e^{i\left[\phi(4M\omega_1)-\phi(4M\omega_2)\right]}e^{i\Delta\omega\bar{v}_0}}{\sqrt{e^{8M\omega_1\pi}-1}\sqrt{e^{8M\omega_2\pi}-1}}\frac{1}{2\pi}\int_{t_0}^{+\infty} dte^{it\Delta\omega},\label{eq:rho_QS_h_cero1}
\end{equation}
with
$t_0=4M\left[\frac{\bar{\omega}}{\Delta\omega}\ln\left(\frac{\omega_2}{\omega_1}\right)+\ln\left(4M_0\sqrt{\omega_1\omega_2}\right)\right]$. This
density matrix is a distribution in $\omega_1-\omega_2$ whose diagonal
yields a divergent term proportional to the number of
particles. The divergence stems from assuming a basis of waves
of definite frequency for the scalar field. As we will see later, the result
can be made finite considering wavepackets with a finite spread in
frequency and time. Computing the diagonal terms we have,
\begin{equation}
\left\langle  N_{\omega}^{QS}\right\rangle_{{\rm Naive\, limit}}=\left\langle  \rho_{\omega,\omega}^{QS}\right\rangle_{{\rm Naive\, limit}}=\int dM \left|\varphi\left(M\right)\right|^2\frac{1}{e^{8M\omega\pi}-1}\frac{1}{2\pi}\int_{\hat{t}_0}^{+\infty} dt,\label{eq:N_QS_h_cero}
\end{equation}
with $\hat{t}_0=4M\left[1+\ln\left(4M_0\omega\right)\right]$. The divergence in $\omega_1=\omega_2=\omega$ appears because we have computed the Bogoliubov coefficients for a continuous basis of plane waves at $I^+$,
\[\left\lbrace\phi_\omega(u)=\frac{1}{\sqrt{2\pi\omega}},
e^{-iu\omega},\omega>0\right\rbrace,\]
and therefore we have considered emission for all time. Formally, from here we can compute the total energy emitted as
\begin{equation}
E_{{\rm Naive\, limit}}=\int_0^{\infty}d\omega\hbar\omega\left\langle  N_{\omega}^{QS}\right\rangle_{{\rm Naive\, limit}},\label{eq:energy_h_cero}
\end{equation}
and of course this will give an infinite result. We should compare
this limit, where we consider fluctuations in the quantum states,
with the ordinary Hawking radiation calculation, where only the energies
of the particles emitted are quantized. Alternatively we can compute a bounded density matrix using
a discrete basis of wavepackets,
\[\left\lbrace\Phi_{\omega_j}(u_n)=\frac{1}{\sqrt{\varepsilon}}\int_{j\varepsilon}^{(j+1)\varepsilon}d\omega\phi_\omega(u)e^{iu_n\omega},j\in\mathbb{N},n\in\mathbb{Z}\right\rbrace,\]
centered around time ($u_n=n\frac{2\pi}{\varepsilon}$) and frequency
($\omega_j=\left[j+1/2\right]\varepsilon$), with
$\varepsilon<<\omega_j$ a narrow frequency window. With this, the
density matrix for the wavepackets becomes, 
\begin{equation}
\left\langle  \rho_{\omega_j,\omega_k}^{QS}\right\rangle_{{\rm Naive\, limit}}(u_n)=\frac{1}{\varepsilon}\int_{j\varepsilon}^{(j+1)\varepsilon}\int_{k\varepsilon}^{(k+1)\varepsilon}d\omega_1 d\omega_2 e^{-iu_n\Delta\omega}\left\langle  \rho_{\omega_1\omega_2}^{QS}\right\rangle_{{\rm Naive\, limit}}
\end{equation}
and the rate of emission of particles is,
\[\left\langle N_{\omega_j}^{QS}\right\rangle_{{\rm Naive\,
      limit}}(u_n)=\left\langle
    \rho_{\omega_j,\omega_j}^{QS}\right\rangle_{{\rm Naive\,
      limit}}(u_n).\] 
Finally this leads to
$E_{{\rm Naive\, limit}}(u_n)$, the power emitted at time $u_n$,
through integral (\ref{eq:energy_h_cero}). This way of computing the
energy allows us to deal with finite quantities and also shows us the
role of the frequency $\omega'$ which appears in the
calculation of the Bogoliubov coefficients and determines their
evolution in terms of the physical time
($u_n$) at $I$.

To do the explicit calculation we need an expression for (\ref{eq:rho_QS_h_cero1}) that we can handle when
$\omega_1\sim\omega_2$.
Approximating the integrand to the lowest order in  $\Delta\omega/\bar{\omega}$ both
in amplitude and in phase in all the factors involved, we get \cite{Skirzewski:2018lql},
\begin{equation}
\left\langle  \rho_{\omega_1\omega_2}^{QS}\right\rangle_{{\rm Naive\, limit}}\sim\int dM \left|\varphi(M)\right|^2e^{i\sigma'(M)\Delta\omega\hbar}\frac{e^{-i4M\Delta\omega\Re\left[\psi^{(0)}(i4M\bar{\omega})\right]}e^{i\Delta\omega\bar{v}_0}}{ e^{8M\bar{\omega}\pi}-1}\frac{1}{2\pi}\int_{\tilde{t}_0}^{+\infty} dte^{it\Delta\omega},\label{eq:rho_QS_aprox2}
\end{equation}
where the change of variable variable (\ref{t}) from $\omega'$ to $t$ becomes
\begin{equation}
t=4M\left[1+\ln\left(4M_0\left[\omega'+\bar{\omega}\right]\right)\right]\label{taprox}
\end{equation}
and therefore $\tilde{t}_0=4M\left[1+\ln\left(4M_0\bar{\omega}\right)\right].$

In the previous expression we have expanded the phase of the function
$\Gamma$ of equation (\ref{eq:gamma_ix1}) as,
\[\phi(x)=\sum_{n=0}^{\infty}\phi^{(n)}(x_0)\frac{(x-x_0)^n}{n!},\]
where $\phi^{(n)}=\Im\left[i^n\psi^{(n-1)}(ix)\right]$ and
$\psi^{(n)}$ is the polygamma function of order $n$. We have also
assumed that the wavefunction in the mass representation takes the form,
\[\varphi(M)=\left|\varphi(M)\right|e^{i\sigma(M)},\]
with $\sigma$ a smooth function and we have expanded $\varphi$
assuming that  $M>>\hbar\Delta\omega$. It is convenient to reorder the
expression (\ref{eq:rho_QS_aprox2}) in the following way,
\[\left\langle\rho_{\omega_1\omega_2}^{QS}\right\rangle_{{\rm Naive\, limit}}\sim\int dM \frac{\left|\varphi(M)\right|^2}{ e^{8M\bar{\omega}\pi}-1}\frac{1}{2\pi}\int_{T_0(\bar{\omega})}^{+\infty}dt e^{i\Delta\omega t},\]
absorbing the phase in $t$ such that,
\begin{equation}
t\to t+\sigma'(M)\hbar-4M\Re\left[\psi^{(0)}(i4M\bar{\omega})\right]+\bar{v}_0,\label{taprox_phase}
\end{equation}
and therefore,
\begin{equation}
T_0(\bar{\omega})=\sigma'(M)\hbar-4M\Re\left[\psi^{(0)}(i4M\bar{\omega})\right]+\bar{v}_0+4M+4M\ln\left(4M_0\bar{\omega}\right).
\end{equation}

We are now in position to incorporate the wavepackets by computing,
\[\langle N_{\omega_j}^{QS}\rangle_{{\rm Naive\, limit}}(u_n)=
\frac{1}{\epsilon}{\int\int_{j\epsilon}^{\left(j+1\right)\epsilon}}d\omega_{1}d\omega_{2}
e^{-u_n\Delta\omega i}\left\langle\rho_{\omega_1\omega_2}^{QS}\right\rangle_{{\rm Naive\, limit}}\sim\]
\[
\sim\int dM \frac{\left|\varphi(M)\right|^2}{e^{8M\omega_{j}\pi}-1}\frac{1}{2\pi\epsilon}
{\int_{-\epsilon}^{\epsilon}}d\left(\Delta\omega\right)
{\int_{\omega_{j}-\frac{\epsilon-\left|\Delta\omega\right|}{2}}^{\omega_{j}+\frac{\epsilon-\left|\Delta\omega\right|}{2}}}\int_{T_0(\bar{\omega})}^{+\infty}dte^{-\left(u_n-t\right)\Delta\omega i}d\bar{\omega}\sim
\]
\[
\sim\int dM \frac{\left|\varphi(M)\right|^2}{e^{8M\omega_{j}\pi}-1}\frac{1}{2\pi}\int_{T_0(\omega_j)}^{+\infty}dt
{\int_{-\epsilon}^{\epsilon}}d\left(\Delta\omega\right)
\frac{\epsilon-\left|\Delta\omega\right|}{\epsilon}e^{-\left(u_n-t\right)\Delta\omega i}=
\]
\begin{equation}=\int dM
  \frac{\left|\varphi(M)\right|^2}{e^{8M\omega_{j}\pi}-1}\frac{\epsilon}{2\pi}\int_{T_0(\omega_j)}^{+\infty}dt\,
  {\rm sinc}^2\left[\frac{\epsilon}{2}(u_n-t)\right]\equiv \int dM\frac{\left|\varphi(M)\right|^2}{e^{8M\omega_{j}\pi}-1}S\left[u_n-T_0(\omega_j)\right],\label{packets12} \end{equation}
where ${\rm sinc}(x)=\sin(x)/x$ is the cardinal sine function and integrates to a function $S$ which is a smooth version of the Heaviside function. 
This expression represents a superposition of thermal radiation that
starts at  time $u_n=T_0(\omega_j)$ and continues to be emitted for later
times. The expression for the emitted power is 
\begin{equation}
E_{{\rm Naive\, limit}}(u_n)\sim\int dM \left|\varphi(M)\right|^2\int_0^{\infty}d\omega \frac{\hbar\omega}{e^{8M\omega\pi}-1}S\left[u_n-T_0(\omega)\right].\label{24}
\end{equation}
Although this is a computation for a quantum shell, by choosing a
state with small uncertainty in the mass (and taking the naive limit
$\hbar\to 0$ in the Bogoliubov coefficients), we are effectively
obtaining the classical limit and therefore the final result coincides
with the usual one quoted for ordinary Hawking radiation for a
classical collapsing shell. 

\section{Corrections to the Bogoliubov coefficients}

To develop a better approximation, we will evaluate
$\beta_{\omega \omega'}^{QS}$ without taking the limit $\hbar\to 0$ in
the integral (\ref{eq:beta_varx}). The latter expression depends on $\hbar$ through
$\Delta_\omega(x)$, $\delta_\omega$ and $m(\omega',\bar{\omega})$.  We
will show  that the approximation described in the previous section
fails to capture important properties of the radiation, in particular
its non-thermal aspects.  To see this, it is convenient to examine
the region $x=0$ of the integral making the change of variable
$y=\ln(x)$. Given that the dependence in $\hbar$ is in the function
$\bar{x}_\omega(x)$ we redefine,
\begin{equation}
\ln\left[\bar{x}_\omega(x)\right]=\ln\left({\rm li}^{-1}({\rm li}(x)-\delta_\omega)\right)={\rm Ei}^{-1}\left({\rm Ei}(y)-\delta_\omega\right)\equiv y_\omega(y), \label{eq:y_w(y)}
\end{equation} 
with ${\rm Ei}$ the exponential integral. 
In figure 2 we show this function
and the approximations to it we will later use. Notice that
$y_\omega[y]\to y$ for large values of $y$. The function $y_\omega(y)$
involves the exponential integral and is not straightforward to
integrate. This will require the use of approximations. 

\begin{figure}%
\subfloat[Function $y_\omega(y)$ and its asymptotes.]{\includegraphics[scale=0.55]{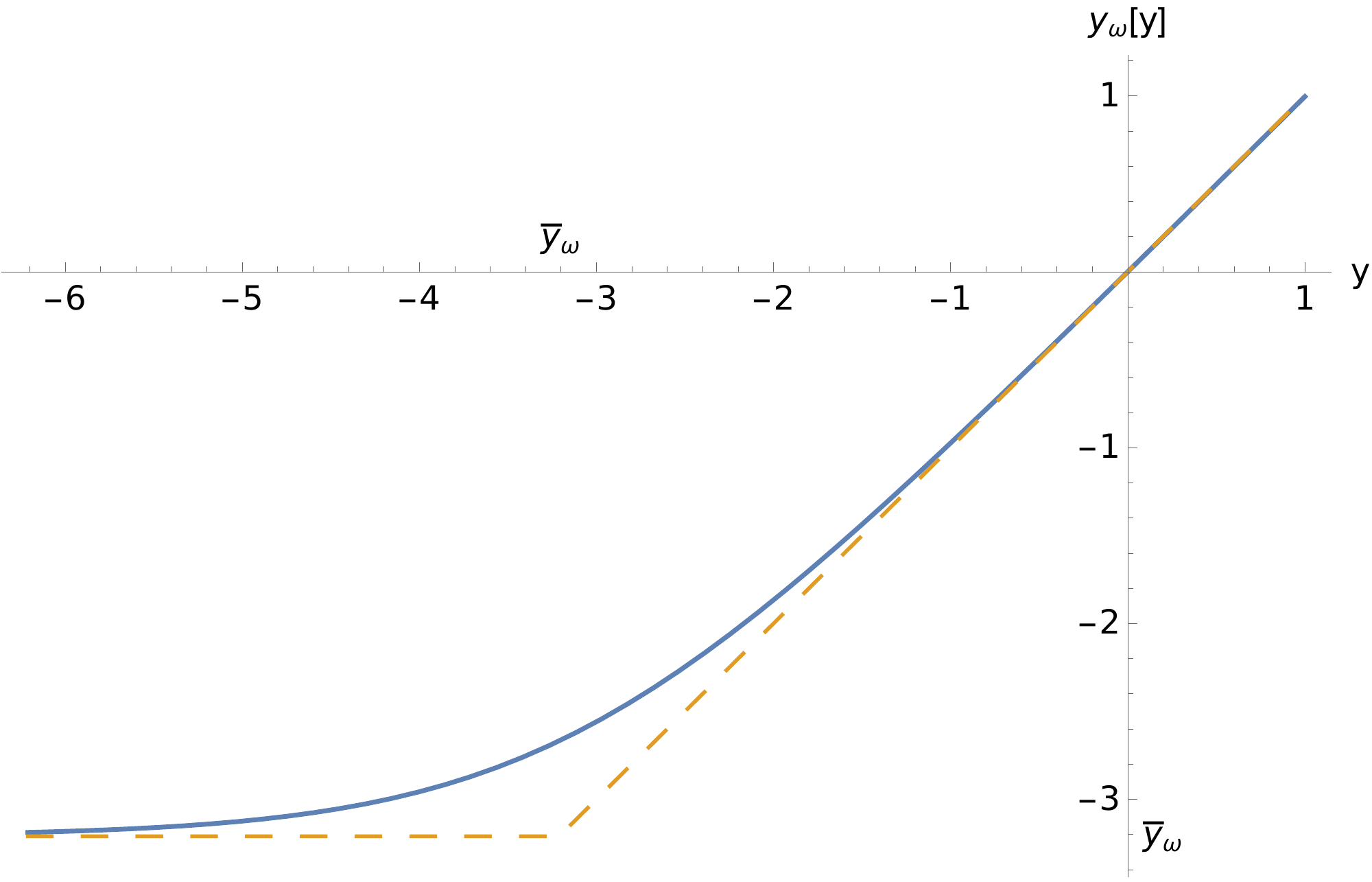}\label{fig:y_omega_aprox_A}}\qquad
\subfloat[Approximation for $\exp\left(y_{\omega}(y)\right)$ (dotted
line) and the exact value.]{\includegraphics[scale=0.55]{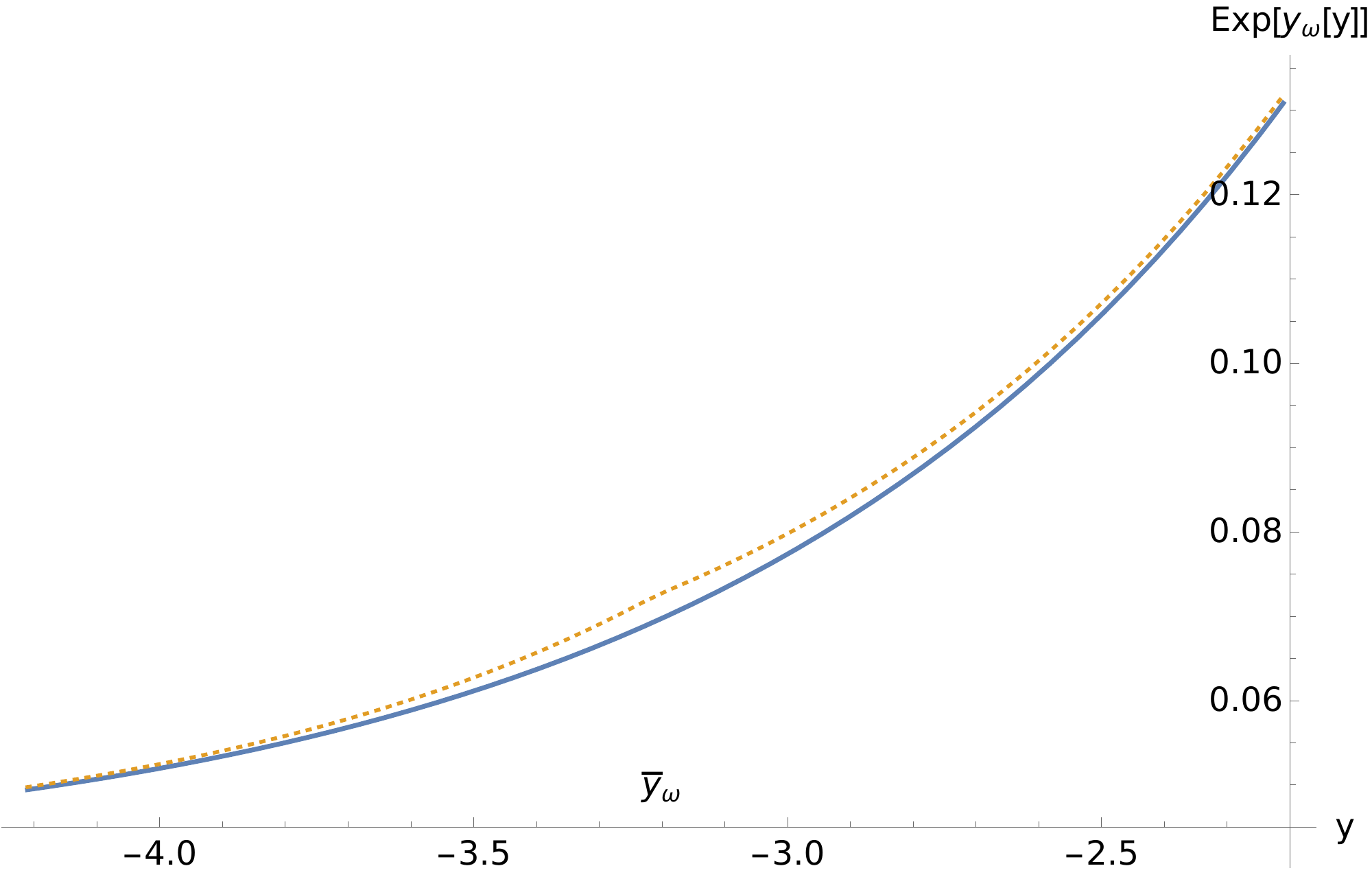}}\qquad
\subfloat[Approximation for
$\sqrt{\left|\frac{y_\omega(y)}{y}\right|}$ (dotted line) and the
exact value.]{\includegraphics[scale=0.55]{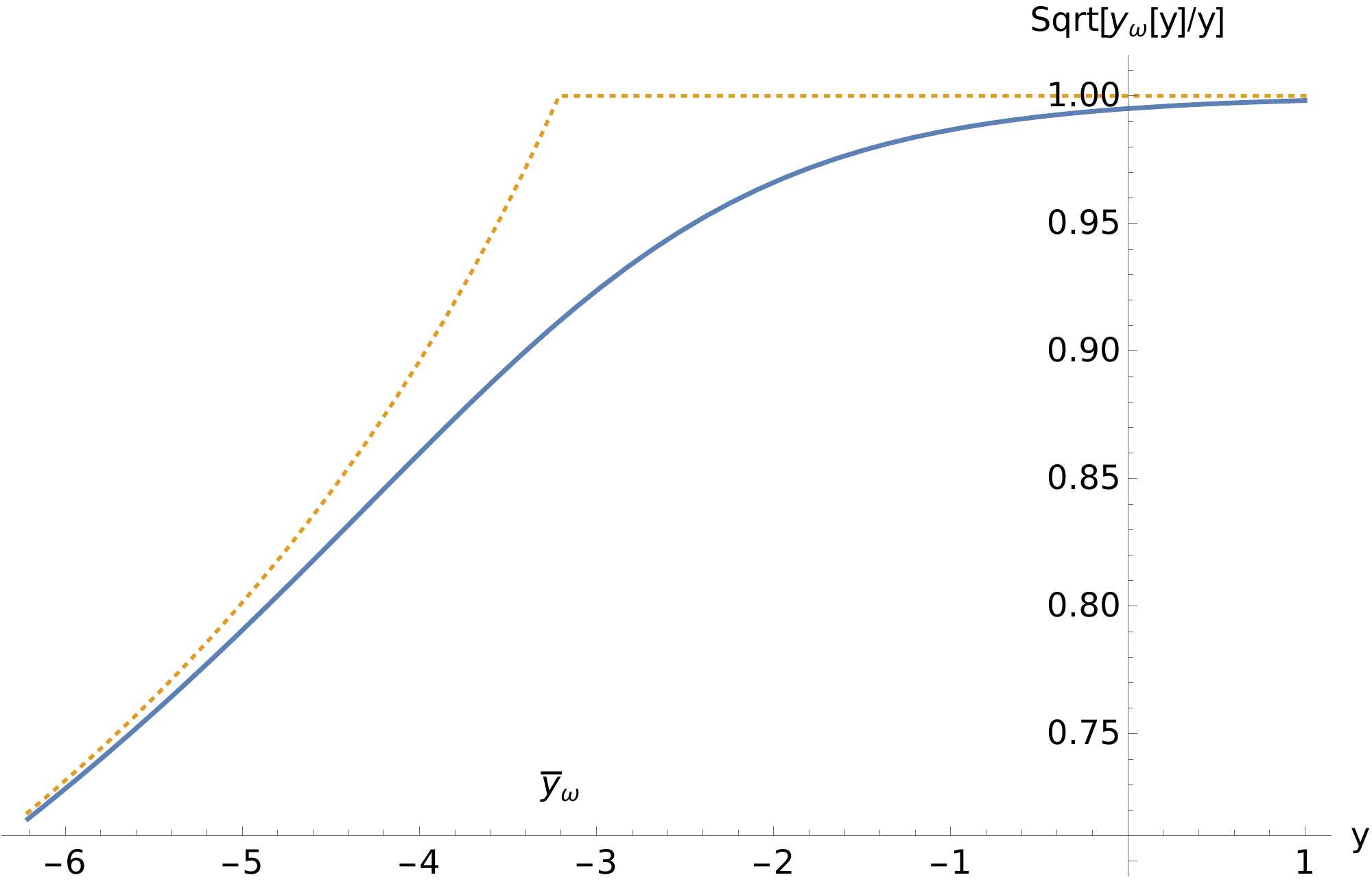}}
\caption{Approximations to the function $y_\omega(y)$ and derived
  functions including the regions at both sides of
  $y=\bar{y}_\omega$. Plots correspond to $\delta_\omega=10^{-2}$ in
  order to have visible departures from the exact functions.}
\label{fig:y_omega_aprox}
\end{figure}

The effective Bogoliubov coefficients (\ref{eq:beta_varx})  of the
quantum shell can be rewritten in terms of $y$ as,
\begin{eqnarray}
\beta^{QS}_{\omega\omega'}(M,\bar{\omega})&=&-\frac{2M_0e^{-i\left[\omega+\omega'\right]\bar{v}_0}}{\pi}\sqrt{\frac{\omega'}{\omega}}\lim_{\epsilon\to0}
{\int}_{-\infty}^{+\infty}dy\exp\left(y-\epsilon e^y\right)\exp\left(i4M_0\left[\omega'+\omega\right]e^y\right)\nonumber\\
&&\times \sqrt{\frac{\left|y_\omega(y)\right|}{\left|y\right|}} \exp\left({i4\omega m(\omega',\bar{\omega})\frac{\exp(y)-\exp(y_\omega(y))}{\delta_\omega}}\right).\label{eq:beta_vary}
\end{eqnarray}

As we see in figure (\ref{fig:y_omega_aprox_A})
there exists a region of the integral where it is incorrect to take the limit 
$\delta_\omega\to0$ (which is equivalent to $y_\omega=y$) in order to
approximate  $y_\omega(y)$ for small values of $x$, which corresponds
to negative values of $y$. In particular for large negative values
$y_\omega(y)$ becomes constant. As figure (\ref{fig:y_omega_aprox_A}) shows, such constant
\[\bar{y}_\omega\equiv y_\omega(-\infty)={\rm Ei}^{-1}(-\delta_\omega),\]
is a good indicator of the value of $y$ where such departure takes place.

\subsection{Asymptotic Approximation}
To take into account these two zones in the calculation of the integral we can approximate the function (\ref{eq:y_w(y)}) using its asymptotic forms for $y\ll\bar{y}_\omega$ and $y\gg\bar{y}_\omega$. When $y>\bar{y}_\omega$ we have that $|{\rm Ei}(y)|>\delta_\omega$ so we consider the approximations,
\begin{eqnarray}
\exp\left[y_\omega(y>\bar{y}_\omega)\right]\approx\exp[y]-y\delta_\omega\label{eq:approx1},\\
\sqrt{\frac{\left|y_\omega(y>\bar{y}_\omega)\right|}{\left|y\right|}}\approx
  1\label{eq:approx2},
\end{eqnarray}
and we will explain why we keep an order less in the square root 
shortly.  This can be obtained expanding (\ref{eq:y_w(y)}) around ${\rm Ei}(y)$. Analogously, when $y<\bar{y}_\omega$, $|{\rm Ei}(y)|<\delta_\omega$ and we approximate,
\begin{eqnarray}
\exp\left[y_\omega(y<\bar{y}_\omega)\right]\approx\exp[\bar{y}_\omega]+\bar{y}_\omega{\rm
    Ei}(y)\label{eq:approx3},\\
\sqrt{\frac{\left|y_\omega(y<\bar{y}_\omega)\right|}{\left|y\right|}}\approx\sqrt{\frac{\left|\bar{y}_\omega\right|}{\left|y\right|}}.\label{eq:approx4}
\end{eqnarray}
 As we show in figure (\ref{fig:y_omega_aprox}), the two approximations
considered coincide when $y=\bar{y}_\omega$. 

These approximations will be used to represent the phase and modulus
in the integral (\ref{eq:beta_vary}). The integral is more sensitive
to the phase, that is why we keep an additional order of approximation
in the exponential with respect to the one taken in the square roots. 

The point where both approximations agree is also where they give their worst result
so is crucial to place a bound on the error introduced in that region
and to find conditions such that the error is small.
Studying the phase and modulus in (\ref{eq:beta_vary}) and comparing
the approximations with the exact values, the conditions are,
\begin{eqnarray}
\frac{4M_0}{\hbar}\left|m(\omega',\omega)\right|\left|\exp[\bar{y}_\omega]-\bar{y}_\omega\delta_\omega-\exp\left[\bar{y}_{2\omega}\right]\right|\ll
  1,\label{eq:cond1}\\
\frac{\sqrt{\left|\bar{y}_{2\omega}\right|}-\sqrt{\left|\bar{y}_{\omega}\right|}}{\sqrt{\left|\bar{y}_{2\omega}\right|}}\ll
  1,\label{eq:cond2}
\end{eqnarray}
where
$\bar{y}_{2\omega}=y_\omega(y=\bar{y}_\omega)=\rm
Ei\left(-2\delta_\omega\right)$. The first condition is imposed
because the approximation appears in the phase of the integrand and
the second because the approximation appears in its modulus. It is
important to notice that the first condition imposes limits for the
range of $\omega'$ and the other one in the range of $\omega$ for
which the approximation is valid.  We will now use these expressions
to compute $\beta^{QS}_{\omega\omega'}$.

\subsection{Approximate computation of  the effective Bogoliubov coefficients}

Considering approximations (\ref{eq:approx1}, \ref{eq:approx2},
\ref{eq:approx3} and \ref{eq:approx4}) we get the following expression
for the effective Bogoliubov coefficients by breaking up the integral
into the two regions involved,
\[\beta^{QS}_{\omega\omega'}(M,\bar{\omega})\equiv\beta^{QS(+)}_{\omega\omega'}(M,\bar{\omega})+\beta^{QS(-)}_{\omega\omega'}(M,\bar{\omega}),\]
with
\begin{equation}
\beta^{QS(+)}_{\omega\omega'}(M,\bar{\omega}) \sim -\frac{2M_0e^{-i\left[\omega+\omega'\right]\bar{v}_0}}{\pi}\sqrt{\frac{\omega'}{\omega}}\lim_{\epsilon\to0}
{\int}_{\bar{y}_\omega}^{+\infty}dy\exp\left({-(\epsilon-i4M_0\left[\omega'+\omega\right])e^y}\right) \exp\left(\left[1+i4\omega m(\omega',\bar{\omega})\right]y\right),
\label{eq:beta_cero_mas}
\end{equation}
for the region $y>\bar{y}_\omega$ and,
\begin{eqnarray}
\beta^{QS(-)}_{\omega\omega'}(M,\bar{\omega})&\sim&-\frac{2M_0e^{-i\left[\omega+\omega'\right]\bar{v}_0}}{\pi}\sqrt{\frac{\omega'}{\omega}}
{\int}_{-\infty}^{\bar{y}_\omega}dy\exp\left({y}\right) \exp\left(i4M_0\left[\omega'+\omega\right]e^{y}\right)\nonumber\\
&&\times \sqrt{\frac{\left|\bar{y}_\omega\right|}{\left|y\right|}} \exp\left({i4\omega m(\omega',\bar{\omega})\frac{e^y-e^{\bar{y}_\omega}-\bar{y}_\omega {\rm Ei}(y)}{\delta_\omega}}\right),
\label{eq:beta_cero_menos}
\end{eqnarray}
for the region $y<\bar{y}_\omega$. We will now study these expressions
assuming $\delta_\omega\ll 1$. The latter is essentially the energy of
the emitted particle divided by the mass of the black hole, so it is
very well satisfied.  We also need to recall that these expressions
are valid only when conditions (\ref{eq:cond1}) and (\ref{eq:cond2})
are met.

\subsubsection{Study of $\beta^{QS(+)}_{\omega\omega'}$}
We start by computing $\beta^{QS(+)}_{\omega\omega'}$.
The integral in (\ref{eq:beta_cero_mas}) can be computed with the change of variable $t=y-\bar{y}_\omega$. Thus,
\begin{eqnarray}\beta^{QS(+)}_{\omega\omega'}(M,\bar{\omega})&=&-\frac{2M_0\exp\left({-i\left[\omega+\omega'\right]\bar{v}_0}\right)}{\pi}\sqrt{\frac{\omega'}{\omega}}\lim_{\epsilon\to0}\exp\left({\left[1+i4\omega m(\omega',\bar{\omega})\right]\bar{y}_\omega}\right)\nonumber\\
&&\times {\int}_{0}^{+\infty}dt\exp\left({-(\epsilon-i4M_0\left[\omega'+\omega\right])e^{\bar{y}_\omega}e^t}\right)\exp\left({\left[1+i4\omega m(\omega',\bar{\omega})\right]t}\right),
\end{eqnarray}
and carrying out the integral in $t$, 
\begin{eqnarray}\beta^{QS(+)}_{\omega\omega'}(M,\bar{\omega})&=&-\frac{i\exp\left({-i\left[\omega+\omega'\right]\bar{v}_0}\right)}{2\pi}\frac{\exp\left({-2m(\omega',\bar{\omega})\omega\pi}\right)}
{\sqrt{\omega'\omega}}\frac{\omega'}{\omega'+\omega}\exp\left({-i4\omega m(\omega',\bar{\omega})\ln\left(4M_0\left[\omega'+\omega\right]\right)}\right)\nonumber\\
&&\times \Gamma\left[1+i4\omega m(\omega',\bar{\omega}),-i\xi(\omega,\omega',M_0)\right]
\label{eq:beta_mas_cero}
\end{eqnarray}
where $\Gamma\left(a,z\right)$ is the incomplete Gamma function and where,
\[\xi(\omega,\omega',M_0)\equiv 4M_0\left[\omega'+\omega\right]\exp\left( {\bar{y}_\omega}\right).\]

\begin{figure}%
\centering
\subfloat[Modulus.]{\includegraphics[scale=0.55]{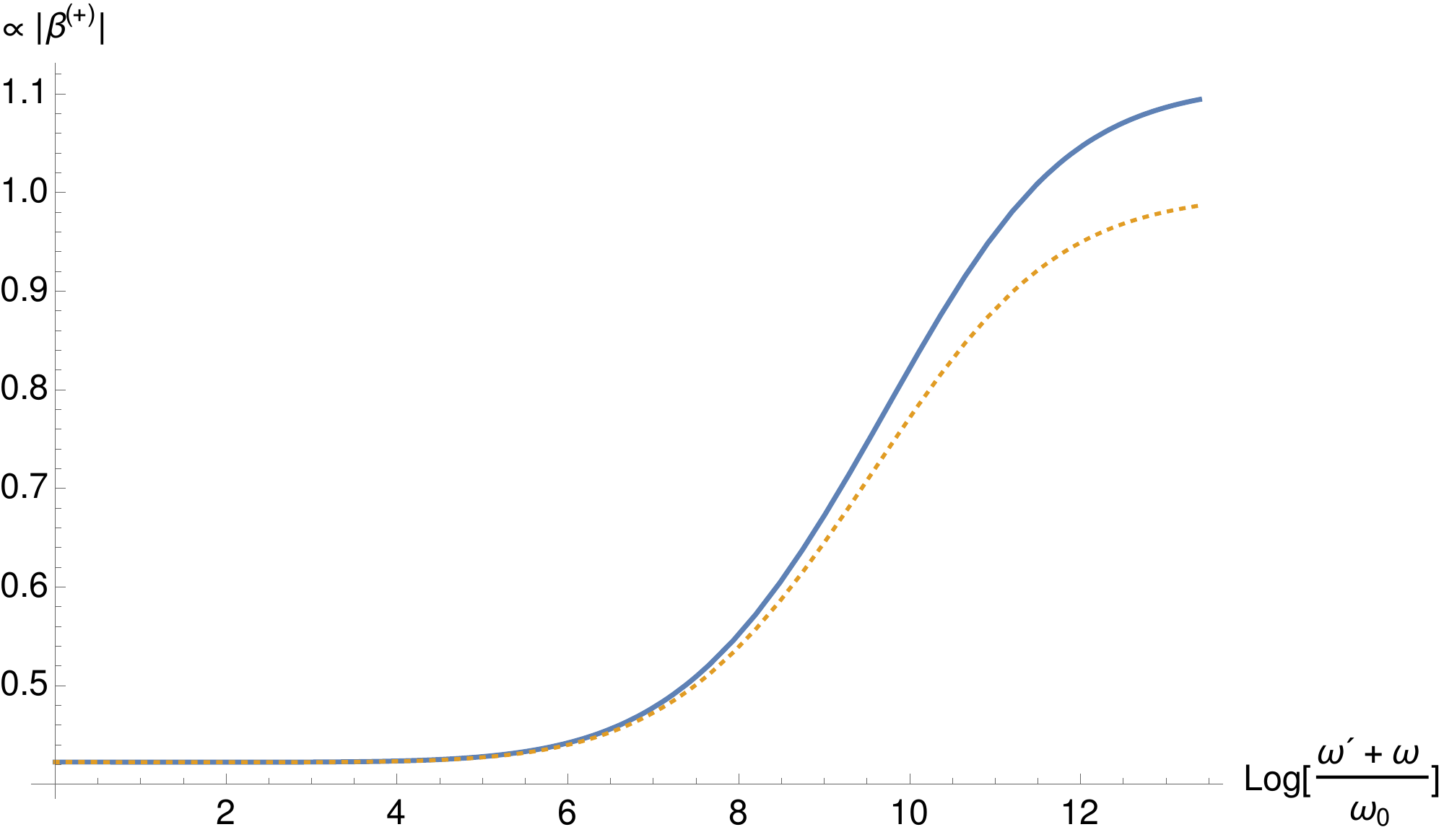}}\qquad
\subfloat[Real part.]{\includegraphics[scale=0.55]{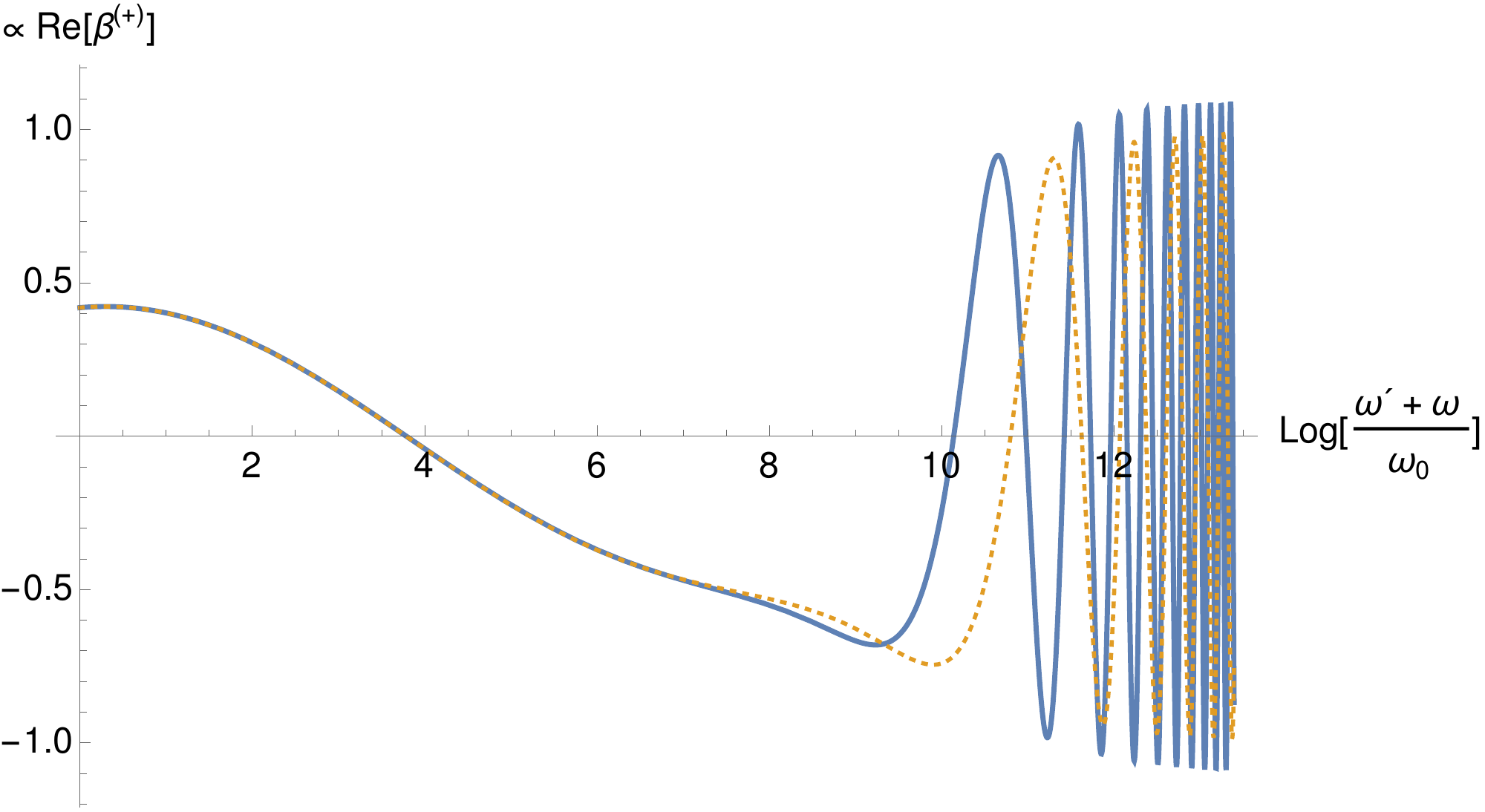}}\qquad
\subfloat[Imaginary part.]{\includegraphics[scale=0.55]{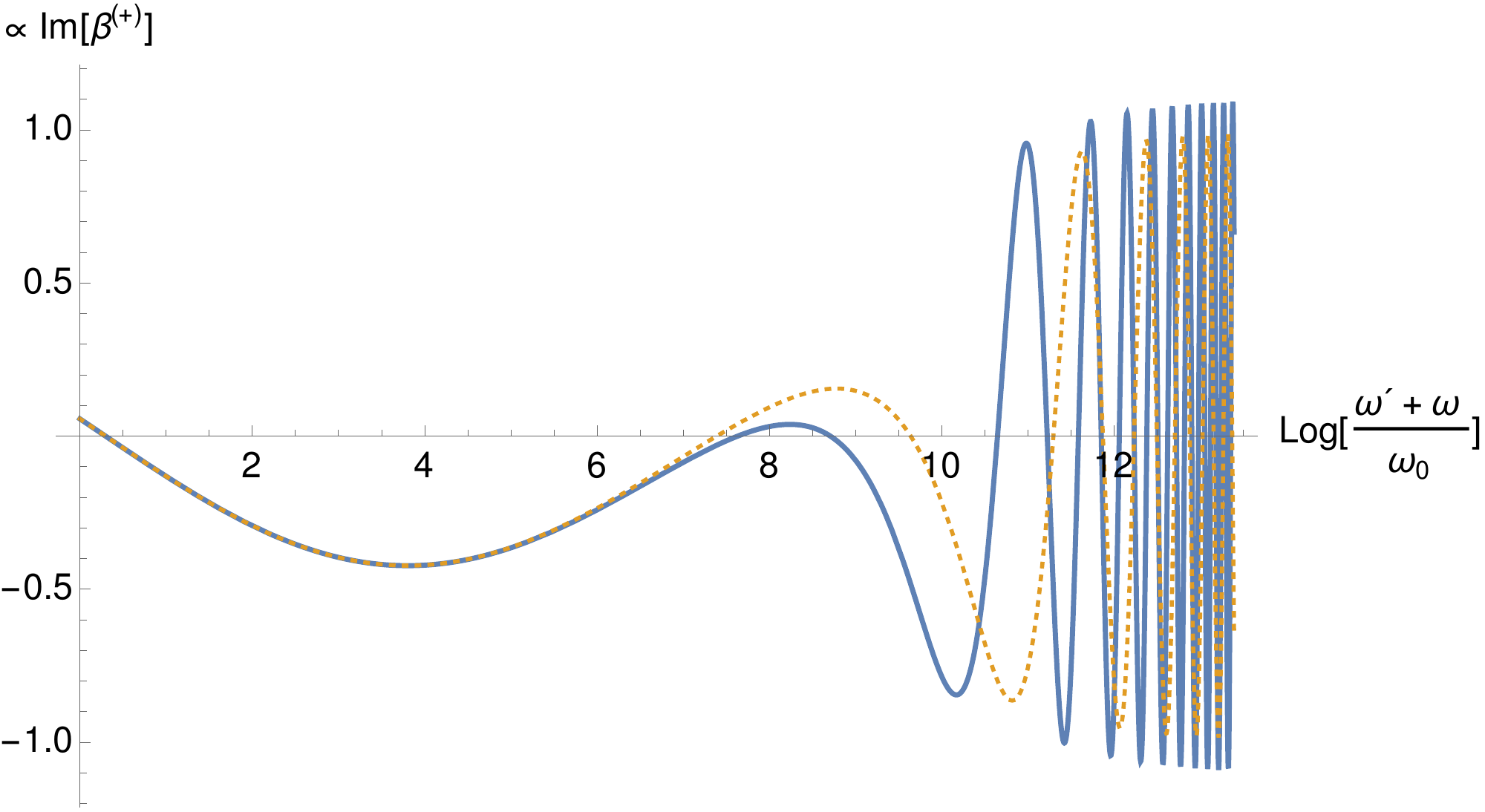}}
\caption{Comparison between $\beta^{QS(+)}_{\omega\omega'}$ (solid
  line) and $\beta^{CS(+)}_{\omega\omega'}$ (dashed line). The latter
  is the integral of the ordinary Hawking calculation in the $t$
  variable starting in $\sqrt{\vert \bar{y}_\omega\vert}$. The plot considers $\omega=\omega_0$ (frequency of maximum emission for Hawking radiation) and $M_0=M$.}
\label{fig:beta_mas_approx}
\end{figure}

Expression (\ref{eq:beta_mas_cero}) reduces to the classical
expression (\ref{eq:bogol_CS}) when $m(\omega',\bar{\omega}) \to M$ and $\xi\to 0$
(equivalent to $\hbar\to 0$, which implies
$\bar{y}_\omega\to-\infty$), but outside this regime, it has a very
different behavior as can be seen in figure
(\ref{fig:beta_mas_approx}), particularly for large values of
$\omega'/\omega_0$. 

Notice that in these plots the pre-factor
\begin{equation}
-\frac{i\exp\left({-i\left[\omega+\omega'\right]\bar{v}_0}\right)}{2\pi}\sqrt{\frac{\omega'}
{\omega}}\frac{1}{\omega'+\omega}
\end{equation}
is omitted and instead of $\omega'$ the plot is made against the
dimensionless variable
$\log\left(\frac{\omega'+\omega}{\omega_0}\right)$ with $\omega_0
\sim 2.82144/(8\pi M_{BH})$ the principal frequency, at which the peak of Hawking
emission occurs. The reasons for the choice of the logarithm will be apparent in the next section.

\subsubsection{Study of $\beta^{QS(-)}_{\omega\omega'}$}

\begin{figure}%
\subfloat[Modulus.]{\includegraphics[scale=0.55]{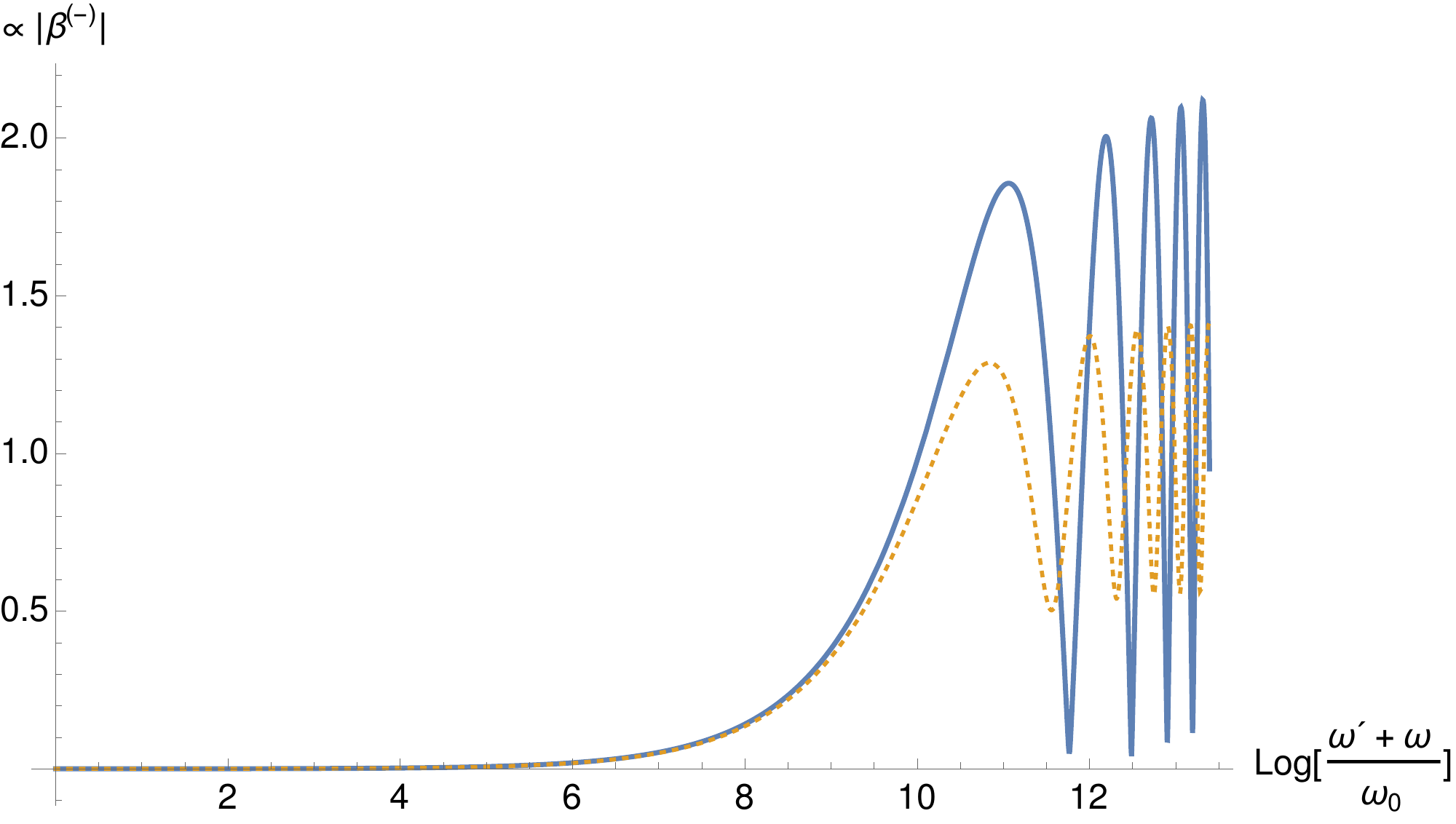}}\qquad
\subfloat[Real part.]{\includegraphics[scale=0.55]{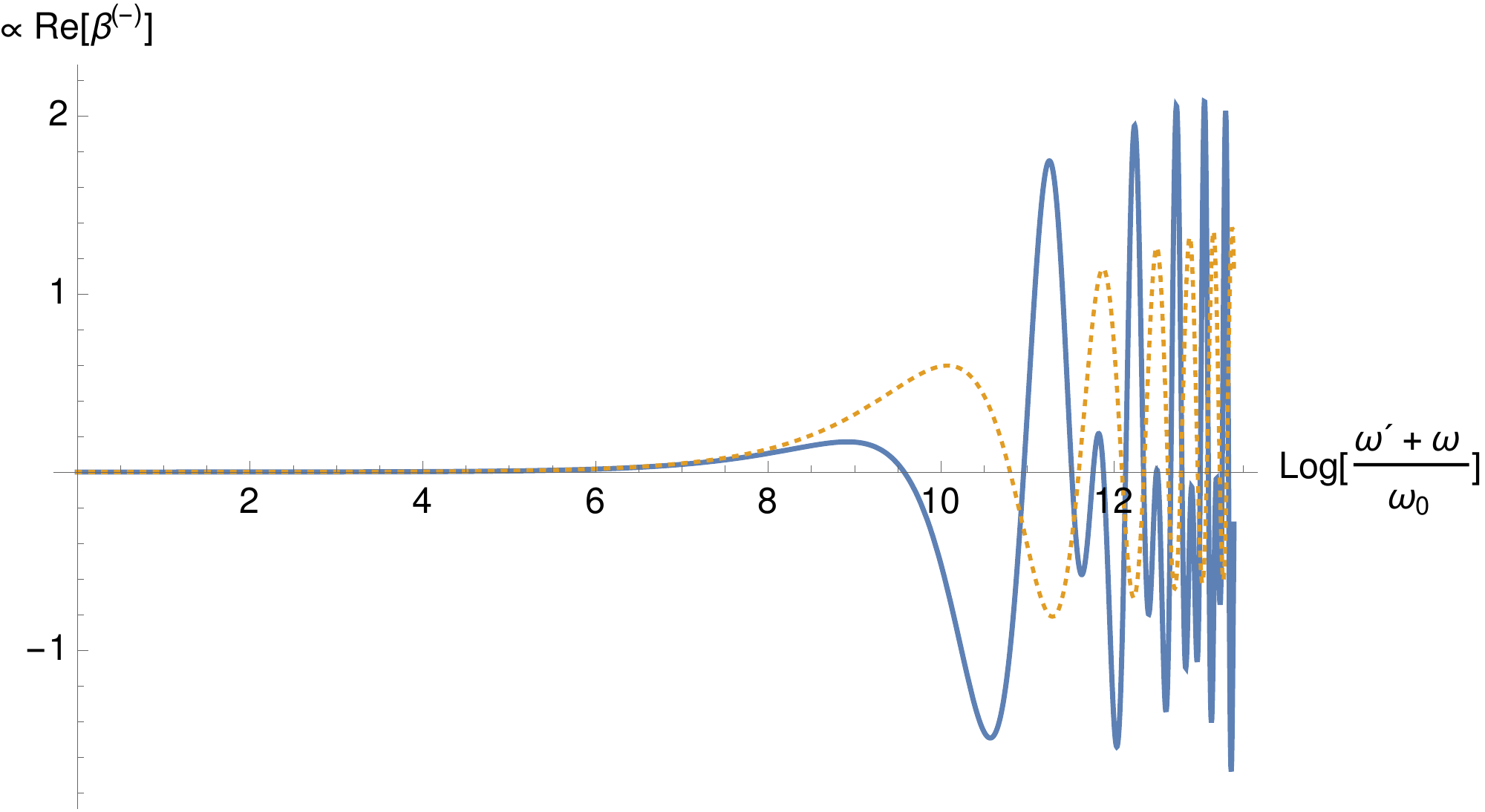}}\qquad
\subfloat[Imaginary part.]{\includegraphics[scale=0.55]{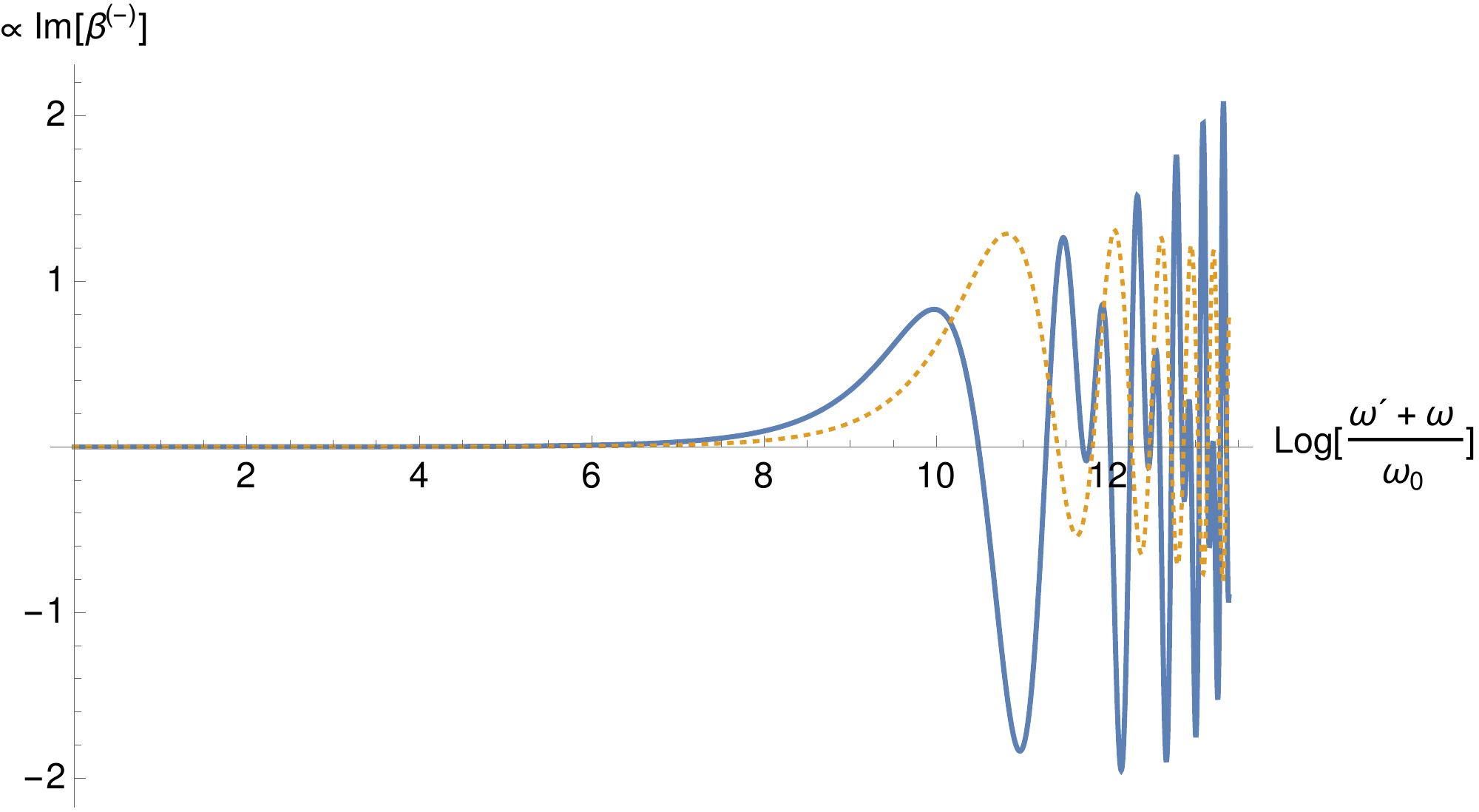}}
\caption{Comparison between $\beta^{QS(-)}_{\omega\omega'}$ (solid
  line) and $\beta^{CS(-)}_{\omega\omega'}$ (dashed line). 
The latter
  is the integral of the ordinary Hawking calculation in the $t$
  variable ending in $\sqrt{\vert \bar{y}_\omega\vert}$. The plot considers $\omega=\omega_0$ (frequency of maximum emission for Hawking radiation) and $M_0=M$.}
\label{fig:beta_menos_approx}
\end{figure}

Let us now concentrate on $\beta^{QS(-)}_{\omega\omega'}$ given by (\ref{eq:beta_cero_menos}). In this case we do not know how to compute the integral in closed 
form. However, unlike $\beta^{QS(+)}_{\omega\omega'}$ this contribution is an integral that converges very fast (due to the real exponentials in the integrand). The change of variable $t=\sqrt{|y|}$ makes it very explicit,
\begin{eqnarray}
\beta^{QS(-)}_{\omega\omega'}(M,\bar{\omega})&\sim&-\frac{2M_0e^{-i\left[\omega+\omega'\right]\bar{v}_0}}{\pi}\sqrt{\frac{\omega'}{\omega}}
{\int}_{\sqrt{|\bar{y}_\omega|}}^{+\infty}2 dt\exp\left({-t^2}\right) \exp\left(i4M_0\left[\omega'+\omega\right]e^{-t^2}\right)\nonumber\\
&&\times \sqrt{\left|\bar{y}_\omega\right|} \exp\left({i4\omega m(\omega',\bar{\omega})\frac{e^{-t^2}-e^{\bar{y}_\omega}-\bar{y}_\omega {\rm Ei}(-t^2)}{\delta_\omega}}\right).
\end{eqnarray} In fact this integral converges absolutely to, $$\sqrt{\pi|\bar{y}_\omega|}{\rm Erfc}(\sqrt{|\bar{y}_\omega|}),$$ where ${\rm Erfc}$ is the complementary error function. In this case the limit $\hbar\to0$ corresponds to $\bar{y}_\omega\to -\infty$ and therefore,
\begin{equation}
\beta^{QS(-)}_{\omega\omega'}(M,\bar{\omega})\to 0,
\end{equation} as is expected for the classical shell. In figure
(\ref{fig:beta_menos_approx}) we see the significant departure of this
expression from the corresponding contribution in the case of the
Hawking radiation for the classical shell, particularly for large
values of $\omega'/\omega_0$. 

\subsubsection{$\beta^{QS}_{\omega\omega'}$ vs $\beta^{CS}_{\omega\omega'}$}

Adding the two contributions previously discussed, we get an
expression for the effective Bogoliubov coefficients
$\beta^{QS}_{\omega\omega'}$ that can be compared with the result
(\ref{eq:bogol_CS}) for the classical shell. In particular, the
modulus $\left|\beta^{QS}_{\omega\omega'}\right|$, evaluated numerically, departs from that of
the classical shell,
\begin{equation}
\left|\beta^{CS}_{\omega\omega'}\right|=\sqrt{\frac{4M}{2\pi}\frac{\omega'}{\omega'+\omega}\frac{1}{\exp(8M\omega\pi)-1}}.
\end{equation}

\begin{figure}%
\centering
\subfloat[Modulus.]{\includegraphics[scale=0.55]{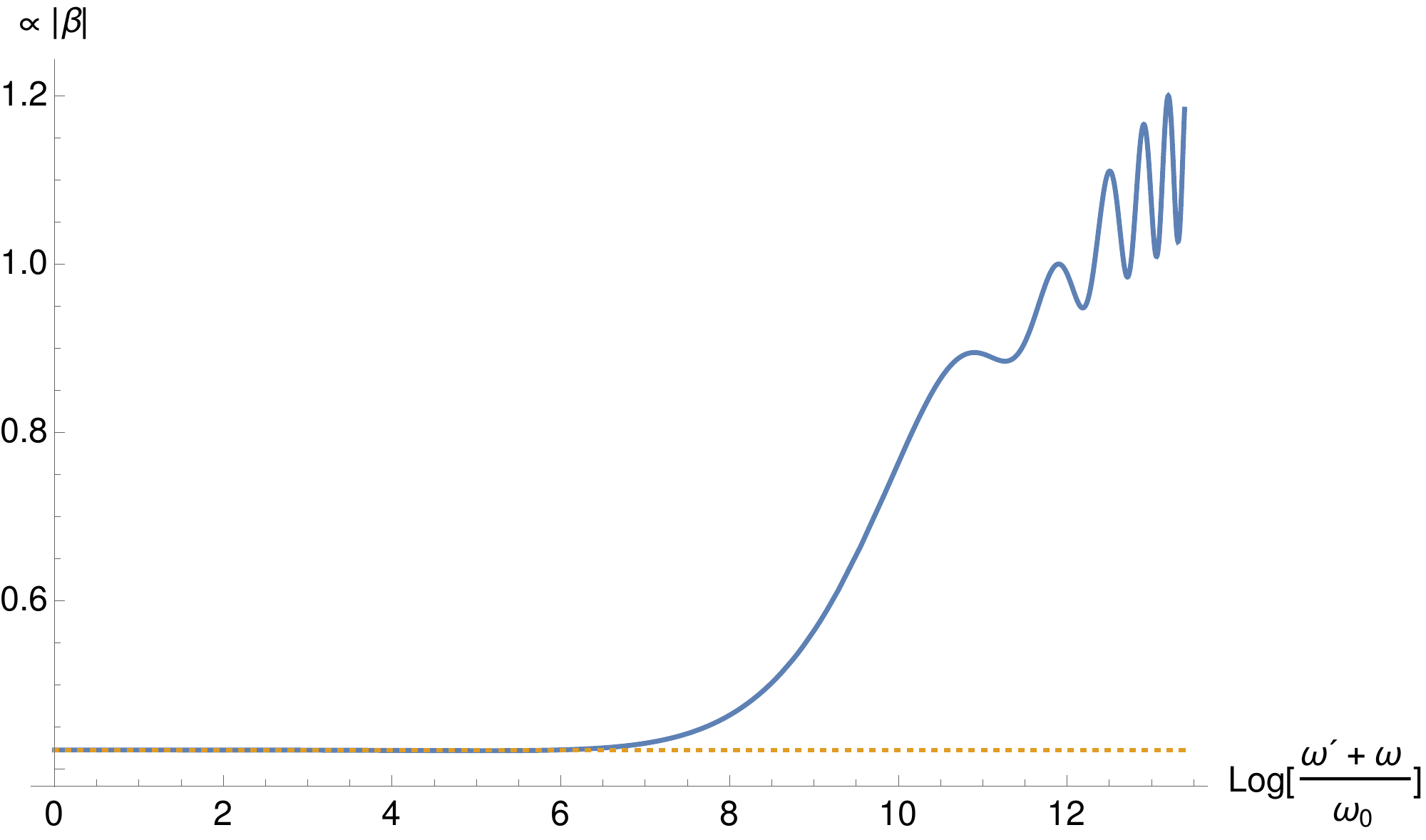}\label{fig:beta_approx_mod}}\qquad
\subfloat[Real part.]{\includegraphics[scale=0.55]{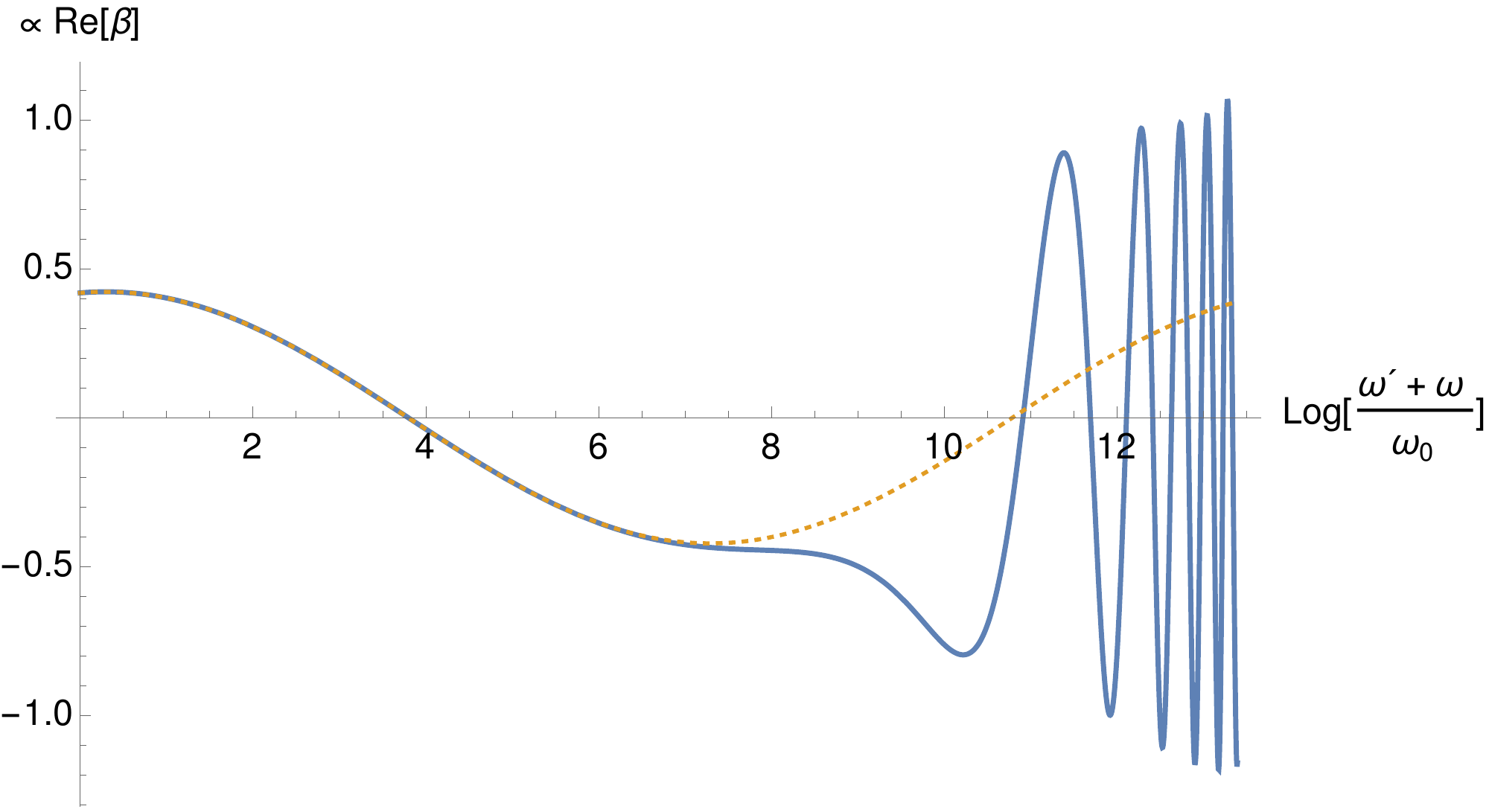}}\qquad
\subfloat[Imaginary part.]{\includegraphics[scale=0.55]{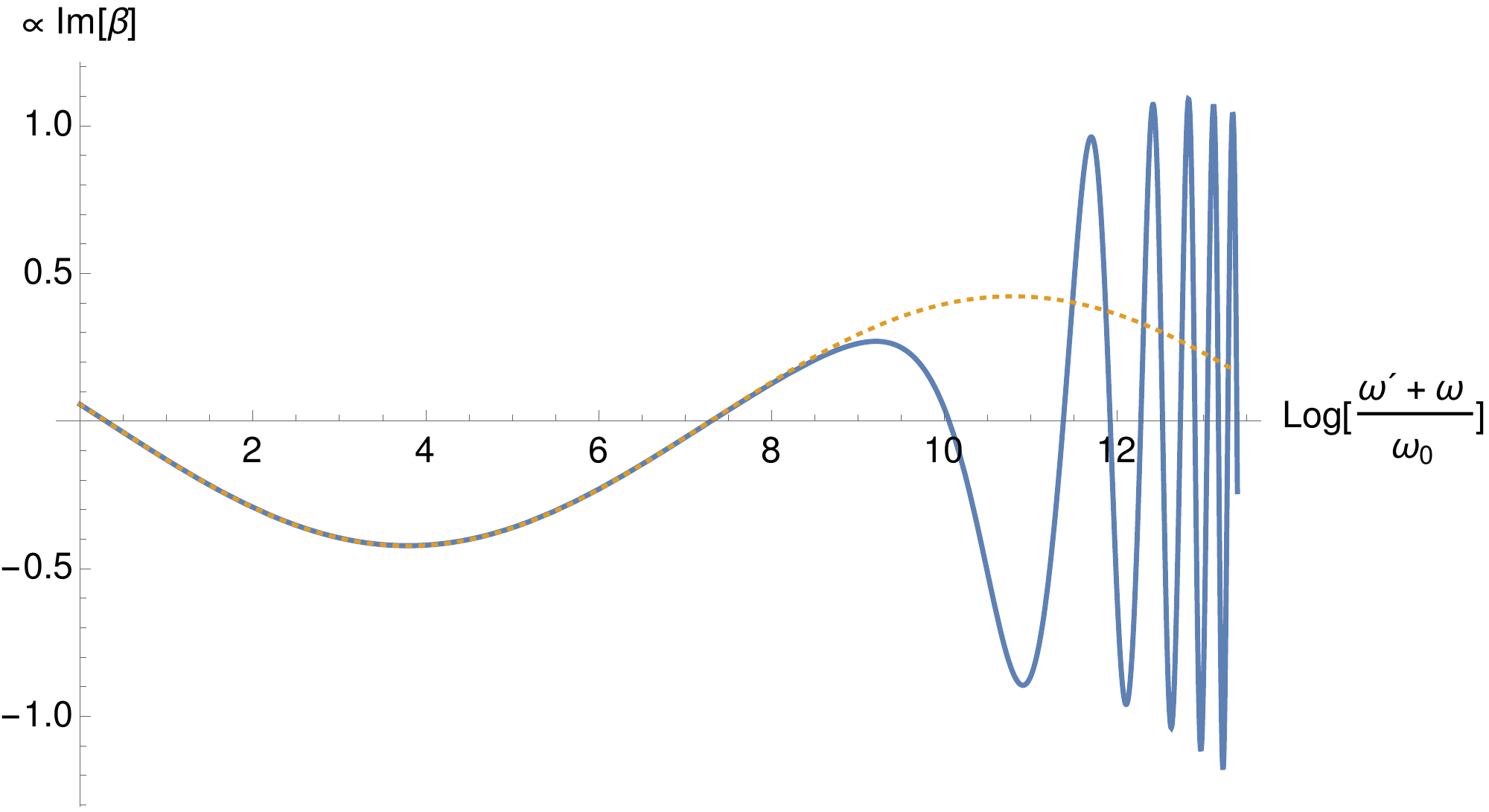}}
\caption{Comparison between $\beta^{QS}_{\omega\omega'}$ (solid line) and $\beta^{CS}_{\omega\omega'}$ (dashed line). The plot considers $\omega=\omega_0$ (frequency of maximum emission for Hawking radiation) and $M_0=M$.}
\label{fig:beta_approx}
\end{figure}

Figure (\ref{fig:beta_approx}) depicts this departure which is most
apparent in (\ref{fig:beta_approx_mod}). The oscillatory behavior at
the end of the plot is not to be trusted. At this
point the frequencies approach the regime where  condition
(\ref{eq:cond1}) is violated and the numerical result is no longer valid.

 In the next section we will discuss the computation of the number of particles and energy emitted for which the Bogoliubov coefficients $\beta$ are the basic ingredient. We will use this numerical result to comment on the departure from the result for the classical shell. 

\section{Number of particles and energy emission based on the previous approximations}

Here we will use the expressions developed in the previous section for
effective Bogoliubov coefficients to compute the number of radiated
particles and the energy emitted. We will compare this result with the
corresponding one that appears in the naive limit discussed in
(\ref{sec:lim_ingenuo}). In the first place we are interested in the
formal integral for continuous frequencies,
\begin{equation}
\left\langle  N_{\omega}^{QS}\right\rangle=\left\langle  \rho_{\omega,\omega}^{QS}\right\rangle=\int dM \left|\varphi\left(M\right)\right|^2\int_0^{\infty}d\omega'\left|\beta^{QS}_{\omega\omega'}(M)\right|^2,\label{eq:rho_QS}
\end{equation}
and finally in the energy
\begin{equation}
E=\int_0^{\omega_{P}}d\omega\hbar\omega\left\langle  N_{\omega}^{QS}\right\rangle=\int dM \left|\varphi\left(M\right)\right|^2\int_0^{\omega_{P}}d\omega\hbar\omega\int_0^{\infty}d\omega'\left|\beta^{QS}_{\omega\omega'}(M)\right|^2\label{eq:energy_QS}. 
\end{equation}
In this expression we have chosen to cutoff the frequency at the
Planck frequency $\hbar \omega_{P}= M_{P}$ with $M_{P}$ the Planck
mass. Higher frequencies would lead to quantum gravity effects and our analysis would not be
valid.

Since these are divergent integrals we can not compare them directly
with (\ref{packets12}) and (\ref{24}).
%(\ref{eq:N_QS_h_cero}) and (\ref{eq:energy_h_cero})
 We could, for example, consider an alternative basis of modes (like
 the wave packets considered in section
 \ref{sec:lim_ingenuo}). However, since we lack an analytic expression
 for the effective Bogoliubov coefficients it would require a
 numerical evaluation that turns out to be very expensive from a
 computational point of view. Instead, we will study the integrand and
 compare it to the one already studied in the naive limit 
$\hbar\to 0$. 

Assuming the wave function of the shell is highly peaked around the
expectation value for the mass $\bar{M}$, we can ignore the
integration in $M$ and focus in the double integral in $\omega$ and
$\omega'$. As we did before, we choose to study these expressions as
functions of $\log\left(\frac{\omega'+\omega}{\omega_0}\right)$ and
$\frac{\omega}{\omega_0}$ where $\omega_0$ is the principal frequency
of emission. These are not only dimensionless but also better related
to the physical variables of the problem (time and energy). Also, we
need to fix the free parameter $M_0$. We chose to set it to $\bar{M}$
because is the usual choice for a classical shell and also because it
makes the conditions $\hbar\omega\ll \bar{M}$ (negligible
back-reaction) and $\delta_\omega=\frac{\hbar\omega}{M_0}\ll 1$
(semi-classical regime) coincide.

\begin{figure}%
  \centering
  {\includegraphics[scale=0.6]{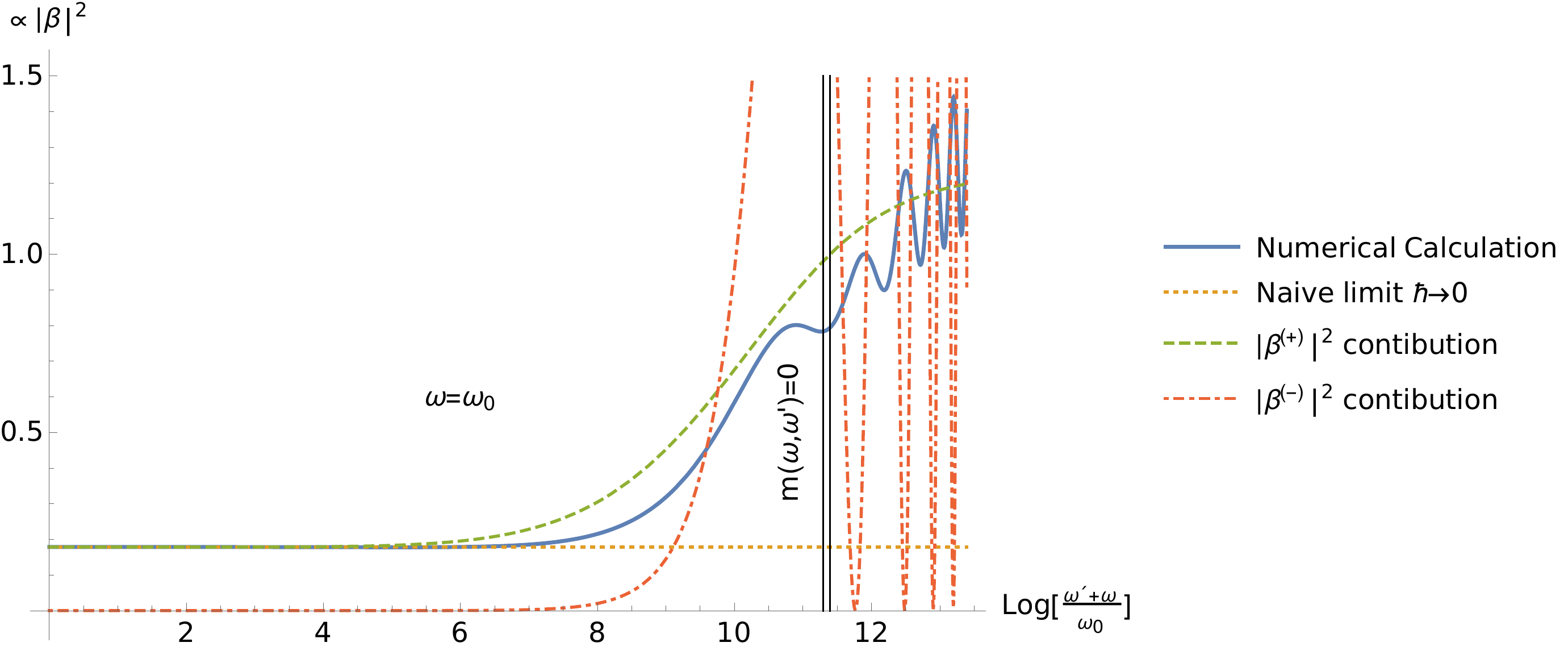}}\caption{Comparison between
    the numerical calculation for
    $\left|\beta^{QS}_{\omega\omega'}\right|^2$, its $(+)$ and $(-)$
    contributions and the naive limit $\hbar\to0$ (corresponding to a
    superposition of Hawking radiation). The plot considers
    $\omega=\omega_0$ and $M_0=M$. The double line represents the
  bound $\log[\frac{M}{\hbar\omega_0}]$ for the independent
  variable. In addition to the plotted squares there are contributions
of crossed terms in the evaluation of the density matrix, not shown.}
\label{fig:beta2_w0}
\end{figure}

In figure (\ref{fig:beta2_w0}) we plot the naive limit
$\left|\beta^{QS}_{\omega\omega'}(M)_{\rm Naive \, limit}\right|^2$, $\left|\beta^{QS}_{\omega\omega'}(M)\right|^2$ and its two contributions $\left|\beta^{QS(+)}_{\omega\omega'}(M)\right|^2$ 
and $\left|\beta^{QS(-)}_{\omega\omega'}(M)\right|^2$. As we did before we omit the prefactor,
\begin{equation}
\frac{1}{(2\pi)^2\omega}\frac{\omega'}{\left(\omega'+\omega\right)^2}.
\end{equation}

We see that $\left|\beta^{QS}_{\omega\omega'}(M)\right|^2$ approaches
the naive limit when $\omega'\to0$ and also that this behavior is
controlled by the $\left|\beta^{QS(+)}_{\omega\omega'}(M)\right|^2$
contribution. This implies the existence of a regime in which the
radiation is thermal and a strong departure when $\omega'$ grows
characterized by an increased rate of emission.  
The condition $m(\omega',\omega)=0$ sets an upper bound for region of
thermal radiation. This is not the only (or best) bound we can find
but is has the advantage of being frequency independent in the
variable $\log\left(\frac{\omega'+\omega}{\omega_0}\right)$,
corresponding to the constant value
$\log\left(\frac{M}{\hbar\omega_0}\right)$.  From these
qualitative analyses we can estimate the amount of energy radiated as
Hawking radiation and also the time when the departure starts.

Assuming the radiation is thermal until $m(\omega',\omega)=0$ then the amount of thermal radiation is,
\begin{eqnarray}
E_{\rm thermal}=\int dM \left|\varphi\left(M\right)\right|^2\int_0^{\omega_{P}}d\omega\hbar\omega\int_0^{m(\omega',\omega)=0}d\omega'\left|\beta^{QS}_{\omega\omega'}(M)_{\hbar=0}\right|^2=\\
=\int dM \left|\varphi\left(M\right)\right|^2\int_0^{\omega_{P}}d\omega\hbar\omega\frac{1}{\exp(8M\omega\pi)-1}\frac{4M}{2\pi}\int_0^{m(\omega',\omega)=0}d\omega'\frac{\omega'}{(\omega'+\omega)^2}=\\
=\int dM \left|\varphi\left(M\right)\right|^2\int_0^{\omega_{P}}d\omega\hbar\omega\frac{1}{\exp(8M\omega\pi)-1}\frac{4M}{2\pi}\left[\log\left(\frac{M}{\hbar\omega}\right)+\frac{\hbar\omega}{M}-1\right]=\\
=\int dM \left|\varphi\left(M\right)\right|^2 M \times I\left(\frac{M}{M_{P}}\right),\end{eqnarray}
where
\begin{equation}
I(x)=x^2\int_0^{M_P/M} ds\frac{s}{\exp(8\pi x^2 s)-1}\frac{s-1-\log(s)}{2\pi}.\end{equation}
This estimate represents less that 0.1\% of the mass of any black hole
with a mass larger than the  Planck mass.

Finally we can make an estimation of the amount of time thermal radiation lasts. Introducing the same basis of wave packets of section (\ref{sec:lim_ingenuo}) and performing the same calculation,  in particular the change of variable (\ref{taprox}) and the subsequent phase absorption (\ref{taprox_phase}), we arrive to the analogous expression for the rate of emitted particles
\begin{equation}
\langle N_{\omega_j}^{QS}\rangle_{\rm thermal}(u_n)=\int dM
  \frac{\left|\varphi(M)\right|^2}{e^{8M\omega_{j}\pi}-1}\frac{\epsilon}{2\pi}\int_{T_0(\omega_j)}^{T_f(\omega_j)}dt\,
  {\rm sinc}^2\left[\frac{\epsilon}{2}(u_n-t)\right], 
\end{equation}
where $T_f(\omega_j)=T_0(\omega_j)+4M\log\left(\frac{M}{\hbar\omega_j}\right)$. This expression represents thermal radiation lasting 
\begin{equation}
\Delta T= 4M\log\left(\frac{M}{\hbar\omega_j}\right),
\end{equation}
which ranges from $\Delta T=4M \log(M/M_{P})$ for $\omega_j\to\omega_{P}$ to $\Delta T=+\infty$ for $\omega_j\to0$. 
\section{Conclusions}
By considering Hawking radiation on the background of a quantum
collapsing null shell, we discovered significant deviations from the
usual Hawking radiation on classical backgrounds. To begin with, 
we obtain thermal radiation that is emitted for a short time (of the
order of the scrambling time, one millisecond for a solar mass black hole), 
insufficient to emit a substantial portion of the mass of the black
hole. After that, a different type of radiation appears with a
non-thermal profile and that can be emitted by a long enough time to
evaporate the black hole. The details of this interval of time depend
on trans-Planckian physics that our model does not capture. For low
frequencies we get thermal radiation that cuts off after a time that
depends on the frequency, giving way to non-thermal radiation of
larger intensity. For lower frequencies the longer the time of
emission of thermal radiation. It always holds that the limit time of emission of thermal
radiation is infinite in the classical limit when $\hbar\to 0$ whereas the emission time
of the non-thermal radiation tends to zero in that limit and the
radiation is always the usual one. It should be noted that the
approximations made are only valid for relatively short periods of
emission. They do not allow to compute correctly the emitted energy
for an arbitrary time of emission. A naive estimate of the evaporation time with the new type of 
radiation found, leads to black holes evaporating considerably faster 
that what traditional Hawking radiation predicts.

The fact that one has non thermal radiation may imply that there does
not exist an information paradox, although a detailed analysis would
be needed of how information could be retrieved, particularly for
collapsing situations that are more realistic than a simple shell.

\section*{Acknowledgment}

We wish to thank Miguel Campiglia and Aureliano Skirzewski
for discussions. This work was supported in part by Grants NSF-PHY-1603630,
NSF-PHY-1903799, funds of the Hearne Institute for Theoretical
Physics, CCT-LSU, Pedeciba and Fondo Clemente Estable
FCE\_1\_2014\_1\_103803.

\end{document}